\documentclass[preprint2,times,tighten]{aastex6}
\pdfoutput=1
\usepackage{amsmath,amstext,apjfonts} 
\usepackage[all]{hypcap}
\usepackage{comment}

\newcommand{\hi}{H\textsc{i}}
\newcommand{\hii}{H\textsc{ii}}
\newcommand{\hidisp}{$\sigma_{\rm \hi}$}
\newcommand{\sigatom}{$\Sigma_{\rm atom}$}
\newcommand{\sigmol}{$\Sigma_{\rm mol}$}
\newcommand{\sigsfr}{$\Sigma_{\rm SFR}$}
\newcommand{\sigstar}{$\Sigma_*$}

\shorttitle{The Origin of Interstellar Turbulence in M33}
\shortauthors{Utomo et al.}

\begin{document}

\title{The Origin of Interstellar Turbulence in M33}
\author{Dyas Utomo\affiliation{1,2}, Leo Blitz\affiliation{2}, and Edith Falgarone\affiliation{3}}
\affil{$^1$Department of Astronomy, The Ohio State University, Columbus, OH 43210, USA; email: utomo.6@osu.edu}
\affil{$^2$Department of Astronomy, University of California, Berkeley, CA 94720, USA}
\affil{$^3${\'E}cole Normale Sup{\'e}rieure, Observatoire de Paris, Universit{\'e} PSL, Sorbonne Universit{\'e}, CNRS, LERMA, F-75005, Paris, France}

\begin{abstract}
We utilize the multi-wavelength data of M33 to study the origin of turbulence in its interstellar medium. We find that the \hi\ turbulent energy surface density inside 8 kpc is $\sim1-3~\times~10^{46}$~erg~pc$^{-2}$, and has no strong dependence on galactocentric radius because of the lack of variation in \hi\ surface density and \hi\ velocity dispersion. Then, we consider the energies injected by supernovae (SNe), the magneto-rotational instability (MRI), and the gravity-driven turbulence from accreted materials as the sources of turbulent energy. For a constant dissipation time of turbulence, the SNe energy can maintain turbulence inside $\sim 4$~kpc radius (equivalent to $\sim 0.5~R_{25}$), while the MRI energy is always smaller than the turbulent energy within 8 kpc radius. However, when we let the dissipation time to be equal to the crossing time of turbulence across the \hi\ scale-height, the SNe energy is enough to maintain turbulence out to 7 kpc radius, and the sum of SNe and MRI energies is able to maintain turbulence out to 8 kpc radius. Due to lack of constraint in the mass accretion rate through the disk of M33, we can not rule out the accretion driven turbulence as a possible source of energy. Furthermore, by resolving individual Giant Molecular Clouds in M33, we also show that the SNe energy can maintain turbulence within individual molecular clouds with $\sim 1\%$ of coupling efficiency. This result strengthens the proposition that stellar feedback is an important source of energy to maintain turbulence in nearby galaxies.
\end{abstract}

\maketitle

\section{Introduction}

The interstellar medium (ISM) is known to be turbulent, from kpc scale of galaxies to sub-pc scale of protoplanetary disk \citep[e.g.,][]{elmegreen04,hennebelle12,maclow04}. At kpc scales, the \hi\ velocity dispersion (\hidisp) in the galactic disk is $\gtrsim 10$ km s$^{-1}$ \citep[e.g.,][]{dickey90,tamburro09,koch18}, larger than the spectral line broadening due to thermal speed \citep[$\sim 8$ km s$^{-1}$ for the warm neutral phase with temperature of $\approx 8 \times 10^3$~K;][]{wolfire95}. This extra kinetic energy that stirs ISM is attributed to turbulence.

The strength of turbulence is important in controlling the star formation activity \citep{krumholz05,federrath12,padoan12,federrath13,salim15}. It is widely known that star formation in nearby galaxies is inefficient. If stars form through gravitational contraction alone, then the typical time-scale is the free-fall time ($\sim10$~Myr), which is much shorter than the molecular gas depletion time in nearby galaxies \citep[$\sim2$~Gyr; e.g.,][]{bigiel08,leroy13,utomo17,utomo18}. Besides magnetic fields, turbulence and stellar feedback (e.g., jets and outflows) have major roles as sources of support to counteract gravity and reduces the star formation rate \citep[][and references therein]{mckee07}. In simulations, by including turbulence, magnetic fields, and stellar feedback step-by-step, the star formation rate is reduced by a factor of $2-3$ with each additional physical ingredient \citep{federrath15}. From observations, \citet{leroy17} showed that the star formation efficiency per free-fall time at the molecular cloud scale in M51 anti-correlates with the velocity dispersion of molecular gas, which is a tracer of the molecular gas turbulent energy.

Turbulence is generated at the driving scale, then it cascades down to smaller scales, and finally dissipates into heat (through viscosity) at a scale comparable to the particle's mean-free-path \citep{frisch95}. The first evidence for turbulent cascades in ISM was reported by \citet{larson81}, who showed that the linewidth of molecular clouds ($\Delta V$) is related to their size ($R$) as $\Delta V \propto R^{0.38}$ \citep[for a compilation of size-line width relation, see][]{hennebelle12}. This empirical relation is similar to the energy cascades of incompressible turbulence predicted by \citet{kolmogorov41}. In a later work, however, \citet{solomon87} revised this scaling relation to be $\Delta V \propto R^{0.5}$ \citep[see also][]{ossenkopf02,heyer04b,roman-duval11}, and hence, the linewidth--size relation is probably just a consequence of the virial equilibrium state of the gas \citep[e.g.,][]{elmegreen89,heyer09,utomo15}. Another alternative is the ISM is highly compressible (because turbulence is supersonic). High-resolution numerical simulations in compressible cold, molecular gas also recovered $\Delta V \propto R^{0.5}$ scaling relation \citep{federrath13}. Observationally, by applying the method from \citet{brunt14}, \citet{orkisz17} found that the selenoidal motion (divergence free) is dominant in the Orion B cloud, but compressive motion is dominant in star-forming regions within Orion~B.

Turbulence dissipates on a short time scale \citep[$\sim 10$~Myr;][]{maclow04}, therefore, the origin of turbulent energy and how it is maintained over the life-time of a galaxy remains problematic. In this paper, we attempt to solve this conundrum by measuring the turbulent energy in atomic and molecular gas in M33, and calculating various possible sources of turbulent energy, such as the magneto-rotational instability (MRI), rotational instability, stellar feedback, and gas accretion from outside the galaxy. For each of them, we use expressions collected from the literature to measure their energy densities, and then, compare it directly to our measured turbulent energy.

M33 is an ideal place to study the interstellar turbulence at least for two reasons. First, the existence of high quality multi-wavelength data (from UV to radio) enables us to compare the turbulent energy with the energy generated from the stellar feedback and MRI. Second, the high resolution data ($\sim 80$ pc of resolution) allow us to study the turbulence down to the scale of molecular clouds. This cloud-scale study is complementary to the kpc-scales study of \citet{tamburro09} and \citet{stilp13}. Compared to previous studies, we provide a more detailed analysis by separating the thermal and turbulent components from the kinetic energy of the gas, and consider the variation of turbulent dissipation time as functions of gas volume density and velocity dispersion of the gas.

This paper is organized as follows. In Sections~\ref{sec:mri} to~\ref{sec:sne}, we review the energy sources that may be able to maintain interstellar turbulence. In Section~\ref{sec:data}, we describe the new and archival data used in our analysis. In Section~\ref{sec:result}, we measure the turbulent energy in azimuthally-averaged binsand also in a cloud-by-cloud basis with an emphasis to explain their energy sources. Lastly, we discuss and summarize our findings in Sections~\ref{sec:discuss} and~\ref{sec:summary}. Throughout this paper, we adopt a distance of $859$ kpc, an inclination of $56^\circ$, and a position angle of $22^\circ.5$ for M33 \citep{gratier10}.

\subsection{Magneto-rotational Instability} \label{sec:mri}

\citet{sellwood99} proposed that turbulence in the outer disk of spirals is driven by the differential rotation of the galaxy under the existence of weak magnetic fields. Their suspicion is based on the fact that \hidisp\ in the outer disk of spirals is roughly constant at $\sim 6$ km s$^{-1}$ \citep{dickey90}, even though stellar winds and supernovae are negligible in that region. They utilized the magneto-rotational instability (MRI) analysis that was first applied to accretion disks by \citet{balbus91}, on galaxy scales, and concluded that a magnetic field strength of $3\mu$G is sufficient for MRI to generate \hidisp\ of 6 km s$^{-1}$. This magnetic field strength is about a factor of 2 smaller than what was measured in the Milky Way by \citet{heiles05}.

The energy per unit area generated by the MRI is (see derivation in Appendix~\ref{app:mri})
\begin{equation} \label{eq:mri_final}
\Sigma_{\rm MRI} \approx 1.1 \times 10^{44} \ {\rm erg \ pc}^{-2} \ \epsilon_{\rm MRI} \ h_{\rm \hi}^2 \ B^2 \ S \ \sigma_{\rm \hi}^{-1},
\end{equation}
where $0 \leq \epsilon_{\rm MRI} \leq 1$ is the coupling efficiency of MRI energy (i.e. the fraction of MRI energy that is deposited as gas turbulence), $h_{\rm \hi}$ is the scale-height of \hi\ gas in units of 100~pc, $B$ is the magnetic field strength in units of $6~\mu$G \citep{tabatabaei08}, $S \equiv |d\Omega/d{\rm ln}R|$ is the shear rate in units of (220~Myr$)^{-1}$, and $\sigma_{\rm \hi}$ is the velocity dispersion of \hi\ gas in units of 10~km~s$^{-1}$. We measure the shear rate in Appendix~\ref{app:kin}, and we record the radial profile of $\Sigma_{\rm MRI}$ in Appendix~\ref{app:table}.

\subsection{Rotational Instability}

In the absence of the Maxwell stress tensor (no MRI), the Newton stress tensor $T_{R \phi} = \langle\rho~u_{GR}~u_{G \phi}\rangle$ can provide the energy input for turbulence from the positive correlation between the radial and azimuthal gravitational velocities \citep[$u_{GR}$ and $u_{G \phi}$, respectively;][]{lynden-bell72}. By adopting the energy input rate from the gravitational instability as $\dot \Sigma_{\rm GI} \approx 1.23 \times 10^{-8}$ erg s$^{-1}$ cm$^{-2}$ \citep{wada02,maclow04}, the energy surface density of the gravitational instability is
\begin{equation}
\Sigma_{\rm GI} = \epsilon_{\rm GI} \ \dot \Sigma_{\rm GI} \ \tau_D \approx 3.6 \times 10^{43} \ {\rm erg \ pc}^{-2} \ \epsilon_{\rm GI},
\end{equation}
where $0 \leq \epsilon_{\rm GI} \leq 1$ is the coupling efficiency of the gravitational instability and $\tau_D \approx 9.8$ Myrs is the dissipation time of turbulence. The calculation above assumes $h_{\rm \hi} = 100 $ pc, $\Sigma_{\rm gas} = 10 \ M_{\odot} \ {\rm pc}^{-2}$, and $S = (220 \ {\rm Myr})^{-1}$. Since this energy is about an order-of-magnitude smaller than MRI, it is not to be considered any further.

\subsection{Gravitational Instability}

A galaxy can accrete cold gas from its intergalactic medium \citep{sancisi08}. Even though the inflowing motion of cold gas is hard to be detected in the inner part of galaxies \citep{wong04}, it has been measured at the outer parts, in the order of $\sim 10$ km s$^{-1}$ \citep{schmidt16}. As the gas settles down, converting its kinetic to potential energy, it drives turbulent motions. The strength of this gravity-driven turbulence was proposed by \citet{krumholz10} as
\begin{equation} \label{eq:grav}
\Sigma_{\rm grav} \approx \frac{3}{4 \pi} \frac{\Omega(r) \ \dot M(r)}{Q \ \eta},
\end{equation}
where $\Omega$ is the angular speed of galaxy, $\dot M$ is the mass accretion rate, $Q$ is the \citet{toomre64} parameter for both stars and gas system, and $\eta$ is a dimensionless number of order unity that measures the turbulent energy dissipation rate per scale height-crossing time. \citet{krumholz16} showed that the gravity-driven turbulence better correlates to the observations of {\it unresolved} local and high-redshift galaxies than a feedback-driven model. We test this model in the resolved observation of M33.

\subsection{Stellar Feedback} \label{sec:sne}

Star formation provides feedback through proto-stellar outflow, stellar winds, and supernovae (SNe). These feedback mechanisms inject energy and momentum to the surrounding ISM. However, SNe energy is orders-of-magnitude higher than the energy from proto-stellar outflows and stellar winds \citep{maclow04}. Therefore, energies from stellar feedback other than SNe are neglected.

We estimate the energy per unit area that needs to be injected by SNe to maintain turbulence in the ISM (i.e. the steady state energy surface density) as $\Sigma_{\rm SNE} = \eta~\epsilon_{\rm SN}~E_{\rm SN}~\tau_D$ \citep{maclow04}, where $\eta$ is the supernovae rate per unit area, $E_{\rm SN}$ is the SN energy, $\epsilon_{\rm SN}$ is the fraction of $E_{\rm SN}$ that goes into turbulence (i.e. coupling efficiency), and $\tau_D$ is the dissipation time, defined as the turbulent crossing time across the turbulent driving scale.

In Appendix~\ref{app:sne}, we estimate $\eta$, $E_{\rm SN}$, and $\tau_D$ to derive the SNe energy surface density as
\begin{equation} \label{eq:sne_final}
\Sigma_{\rm SNE} \approx 2.0 \times 10^{46} \ {\rm erg \ pc}^{-2} \  \epsilon_{SN} \ \left( \frac{\Sigma_{\rm SFR}}{M_{\odot} \ {\rm Gyr}^{-1} \ {\rm pc}^{-2}} \right),
\end{equation}
where $0 \leq \epsilon_{SN} \leq 1$ is the coupling efficiency of SNe energy and \sigsfr\ is the star formation rate surface density. In deriving Equation~\ref{eq:sne_final}, we assume that the momentum injected by a SN in two-phase medium is $2.8 \times 10^5 M_{\odot}$ km s$^{-1}$ \citep{kim15} and $\tau_D$ is constant \citep[9.8 Myr;][]{maclow04}. However, we also consider the SNe energy for $\tau_D$ equal to the crossing time of turbulence across the \hi\ scale-height. We record the radial profile of $\Sigma_{\rm SNE}$ in Appendix~\ref{app:table}.

\section{Data} \label{sec:data}

\begin{figure*}
\centering
\includegraphics[width=\textwidth]{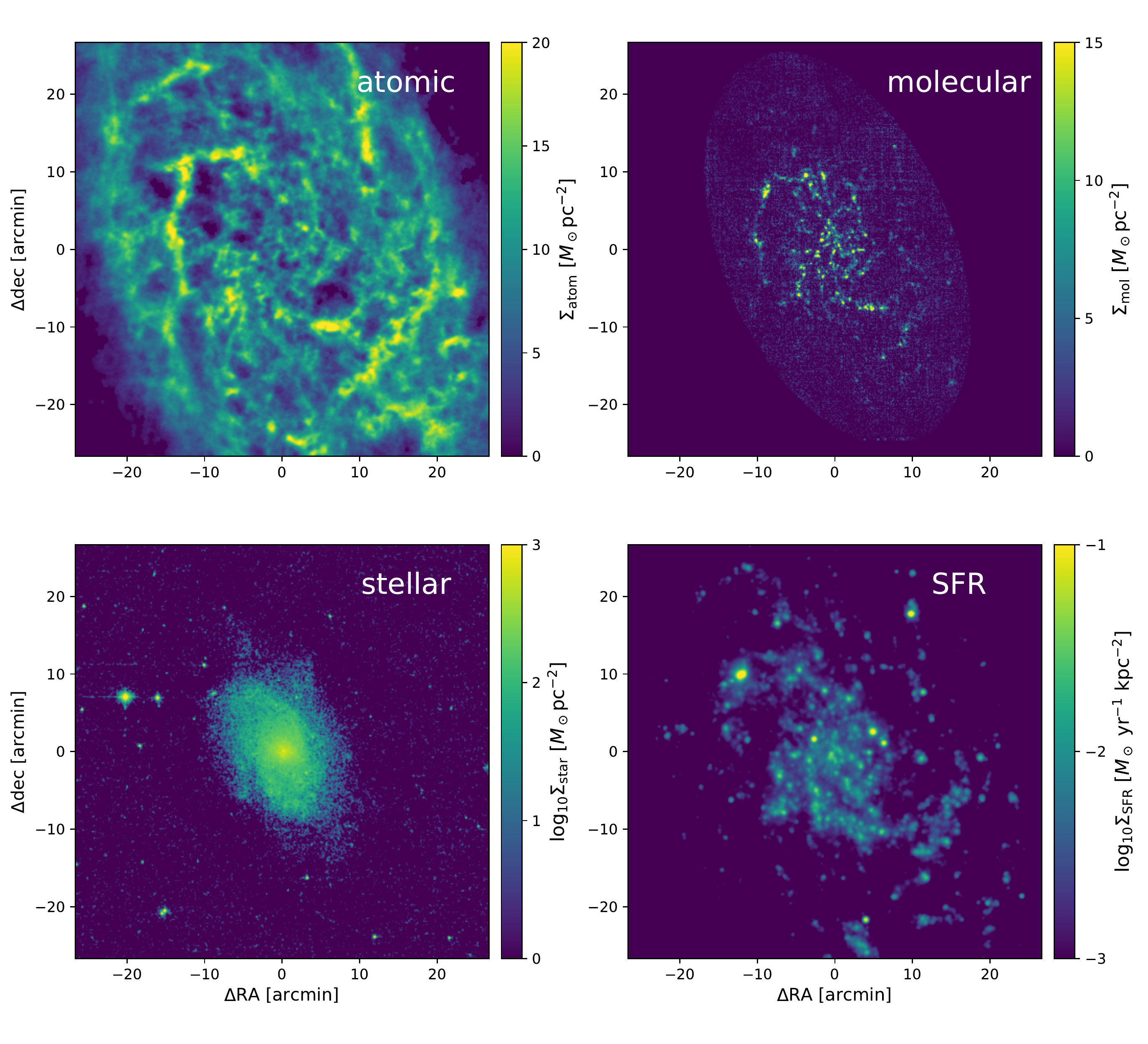}
\caption{The map of the atomic gas surface density from VLA and GBT \citep{koch18}, color coded in $M_\odot$~pc$^{-2}$ (top left), the molecular gas surface density in $M_\odot$ pc$^{-2}$ from IRAM \citep[][top right]{gratier10,druard14}, the stellar surface density in log($M_\odot$ pc$^{-2}$) from 2MASS \citep[][bottom left]{jarrett03}, and the SFR surface density in log($M_\odot$ yr$^{-1}$ kpc$^{-2}$) from {\it GALEX} and {\it Spitzer} \citep[][bottom right]{depaz07,rieke04,gordon05,dale09}.}
\label{fig:gallery}
\end{figure*}

\subsection{Atomic Gas}

The data cube of \hi\ emission is retrieved from \citet{koch18}. This data cube is a combination of new interferometric observations from the Karl G. Jansky Very Large Array (VLA; Project ID 14B-088) and archival single dish observations from the Robert C. Byrd Green Bank Telescope (GBT; 2002 October). This combination ensures that \hi\ emission from both small and large spatial scales is captured. Reduction and analyses of the raw data have been done by \citet{koch18}, resulting in a data cube with 80 pc of linear resolution and 0.2 km s$^{-1}$ of spectral resolution.

\subsubsection{Mass Surface Density}

The \hi\ integrated intensity ($I_{\rm \hi}$) is converted to the \hi\ mass surface density ($\Sigma_{\rm atom}$), assuming \hi\ is optically thin throughout the line-of-sight, via \citep{leroy08}
\begin{equation} \label{eq:mass_atom}
\frac{\Sigma_{\rm atom}}{M_{\odot} \ {\rm pc}^{-2}} = 0.020 \ {\rm cos}(i) \ \frac{I_{\rm \hi}}{{\rm K \ km \ s}^{-1}},
\end{equation}
where $i = 56^{\circ}$ is the inclination of M33 \citep{gratier10}. Equation~\ref{eq:mass_atom} includes a factor of 1.36 to reflect the presence of Helium. The map of \hi\ surface density ($\Sigma_{\rm \hi}$) is shown in Figures~\ref{fig:gallery}. Its radial profile is shown in Figure~\ref{fig:rad_prof} and is recorded in Appendix~\ref{app:table}.

\subsubsection{Velocity Dispersion}

There are two common ways to derive the velocity dispersion (\hidisp): the second-moment and Gaussian fit. However, both methods are susceptible to noise if the signal-to-noise ratio (SNR) is low. Therefore, we average the \hi\ {\it stacked}-spectra within radial bins (100 pc wide) to increase SNR. The \hi\ stacked-spectra are defined as the \hi\ spectra, shifted into a common velocity reference within a radial bin, and then, taking their average \citep[e.g.,][]{ianjamasimanana12,stilp13,koch18}. By applying this method, we can get a high SNR, including the spectral edges (line wings) that are important to the measurements of \hidisp.

There are three choices of the reference velocity to stack the spectra: the centroid velocity (derived from the first moment), the peak velocity (i.e. velocity that corresponds to the highest emission in the spectrum), and the galaxy rotation velocity at the respective galactocentric radius. \citet{koch18} have done a detailed comparison of those choices, and concluded that the peak velocity is less biased. Hence, we also follow their suggestion to use the peak velocity as a reference. Using the centroid or rotation velocities as a reference leads to a larger velocity dispersion. Therefore, we also consider those velocity references as the upper limit of the derived velocity dispersions.

The resulting stacked spectra deviate from a Gaussian because of extra-flux at the line wings. There are multiple interpretations of these line wings, e.g. warm \hi\ component, outflows, turbulent motion, and lagging \hi\ component above the midplane. \citet{koch18} showed that $\sim 9\%$ of the line flux is originated from the lagging \hi\ component above the midplane, i.e. not of turbulent origin. This lagging \hi\ component manifests itself as asymmetric line wings. Since this value is quite small, we assume that the line wings are originated from turbulent motions. However, we also conside excluding that asymmetric component to calculate the lower limit of the velocity dispersions.

To take into account the line wings, we fit each stacked spectrum with double Gaussians (narrow and broad components). Then, we measure the area and dispersion of each Gaussian component. We define the velocity dispersion of the line as the mean of \hidisp, weighted by the Gaussian area, $A$, of each component, i.e.
\begin{equation}
\sigma_{\rm \hi} = \frac{A_{\rm narrow} \sigma_{\rm narrow} + A_{\rm broad} \sigma_{\rm broad}}{A_{\rm narrow} + A_{\rm broad}}.
\end{equation}
The resulting \hidisp\ is almost constant (independent of radius) with values between $\sim 10$ and 13 km s$^{-1}$, which is shown in Figure~\ref{fig:hi_disp} and recorded in Appendix~\ref{app:table}. Velocity dispersion in high resolution observations is usually less affected by beam smearing, except at the center of galaxy. We model this artificial broadening due to beam smearing in Appendix~\ref{app:smear}.

\subsubsection{Kinetic Energy Surface Density}

The atomic gas kinetic energy per unit area is
\begin{equation} \label{eq:turb_energy}
\Sigma_{\rm kin} = \frac{3}{2} \ \Sigma_{\rm HI} \ \sigma_{\rm HI}^2.
\end{equation}
The factor of 3 in Equation~\ref{eq:turb_energy} is included to calculate the 3-dimensional kinetic energy from 1-dimensional velocity dispersion, with an assumption that \hidisp\ is isotropic. We record the radial profile of $\Sigma_{\rm kin}$ in Appendix~\ref{app:table}. Since the kinetic energy of the gas originates from thermal and turbulent motions, we separate those thermal and turbulent motions in $\S$\ref{sec:separate}.

\subsubsection{Scale Height}

We assume that the vertical distribution of \hi\ gas is in dynamical equilibrium, where the weight of \hi\ gas under the influence of the gravitational potential of the galaxy (stellar and gas, excluding dark matter) is balanced by the pressure gradient of the \hi\ gas. Following \citet{ostriker10} and \citet{kim13}, the total pressure in this dynamical equilibrium state is
\begin{equation} \label{eq:pressure}
P_{\rm tot,DE} = f_{\rm d} \frac{\pi G \Sigma^2}{4} \left[(2-f_{\rm d}) + \sqrt{(2-f_{\rm d})^2 + \frac{32 \sigma_{\rm \hi}^2 \rho_\star}{\pi^2 G \Sigma^2}} \ \right],
\end{equation}
where $\Sigma = \Sigma_{\rm \hi} + \Sigma_{\rm mol}$ is the total gas surface density, and we define the diffuse gas fraction as $f_{\rm d} = \Sigma_{\rm \hi}/\Sigma$. Some authors consider the CO emitting gas as diffuse and only gas in the cores is self-gravitating. In this case, $f_{\rm d} \approx 1$ and we underestimate the value $P_{\rm tot, DE}$.

We can not measure the stellar volume density, $\rho_\star$, directly, so we estimate it as $\rho_\star~\approx~\Sigma_\star/(2h_\star)$. We use the flattening ratio $\ell_\star/h_\star = 7.3$ \citep{kregel02} and the stellar length scale $\ell_\star = 2.3$~kpc \citep{vandenbergh91} to calculate $h_\star$.

The vertical hydrostatic pressure is balanced by the {\it one-dimensional} volumetric kinetic energy of \hi\ gas, so that the \hi\ gas scale-height in hydrostatic equilibrium state is
\begin{equation}
\label{eq:height}
h_{\rm \hi} \approx \frac{0.5 \ \Sigma_{\rm \hi} \ \sigma_{\rm \hi}^2}{2 P_{\rm tot,DE}}
\end{equation}
A factor of 2 in the denominator arises because $\rho \approx \Sigma/(2h)$. In a more realistic case, the \hi\ scale-height is time dependent and undergoes oscillation \citep[e.g.,][]{benincasa16}. Hence, Equation~\ref{eq:height} can be interpreted as either a static case or an average over time-scale of $\sim 100$ Myrs.

We show the radial profile of $h_{\rm \hi}$ in the right panel of Figure~\ref{fig:hi_disp} and record it in Appendix~\ref{app:table}. The uncertainties are calculated from the error propagations. The \hi\ scale height increases from $\approx 72$~pc in the center to $\approx 627$~pc at $ 8$~kpc radius, mostly driven by the decrease in the total pressure (Equation~\ref{eq:pressure}).

\begin{figure}
\centering
\includegraphics[width=0.48\textwidth]{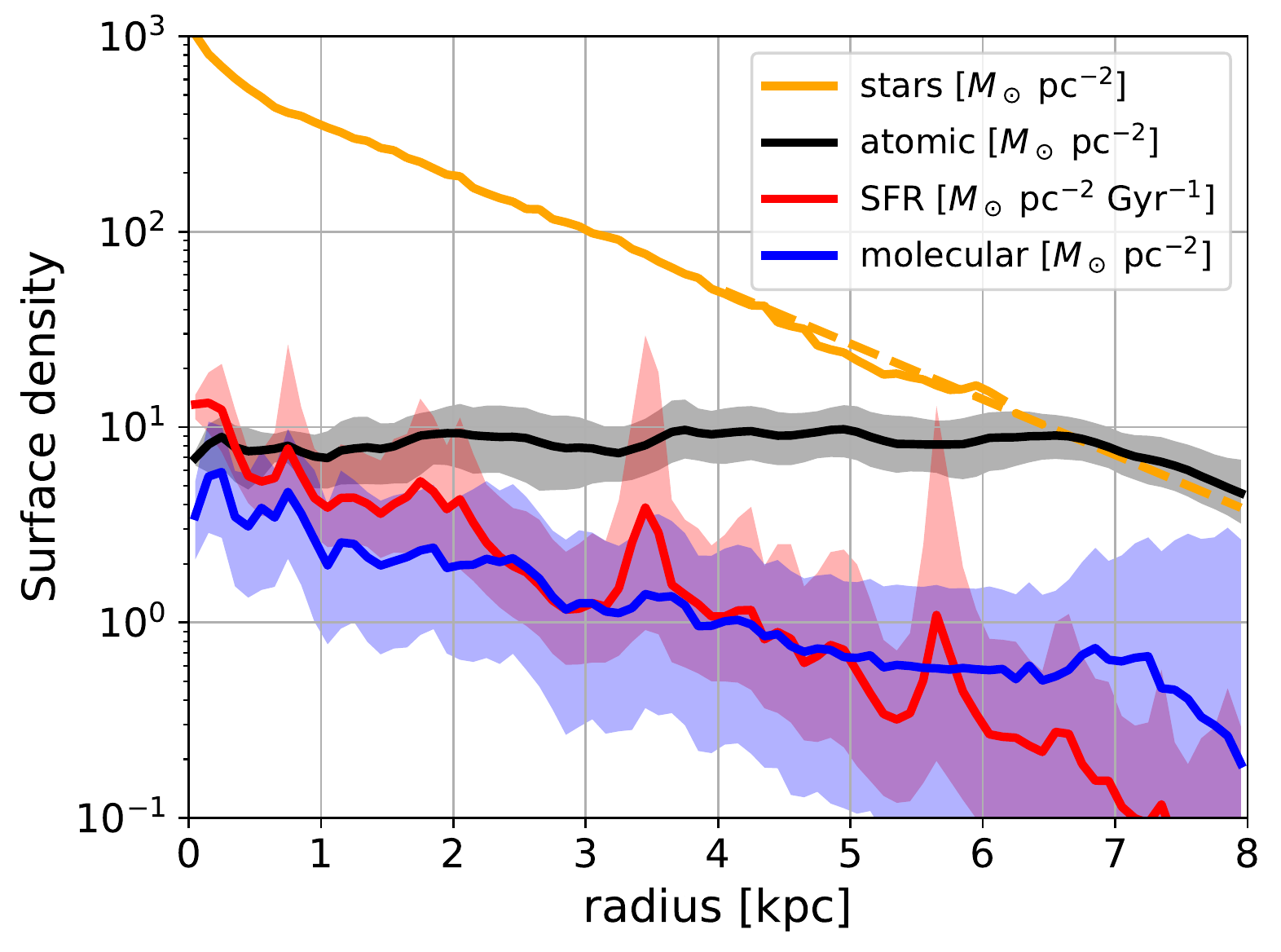}
\caption{The radial profiles of \sigatom\ in units of $M_\odot$ pc$^{-2}$, \sigmol\ in units of $M_\odot$ pc$^{-2}$, \sigstar\ in units of $M_\odot$ pc$^{-2}$, and \sigsfr\ in units of $M_\odot$ Gyr$^{-1}$ pc$^{-2}$ by taking their average values in radial bins with bin width of 100~pc. The dashed line in \sigstar\ profile is the extrapolation from the exponential disk fit. The statistical uncertainty in \sigstar\ is very small. The gas is atomic dominated throughout the galaxy. Two spikes in \sigsfr\ at radius $\sim 3.5$ kpc and $\sim 5.7$ kpc are from NGC604 and IC133, respectively. The typical uncertainties are derived from the median absolute deviation of the respective surface density in each radial bin (i.e. it takes into account variations in the azimuthal direction) and are marked as colored bands.}
\label{fig:rad_prof}
\end{figure}

\begin{figure*}
\centering
\epsscale{1.15}
\plottwo{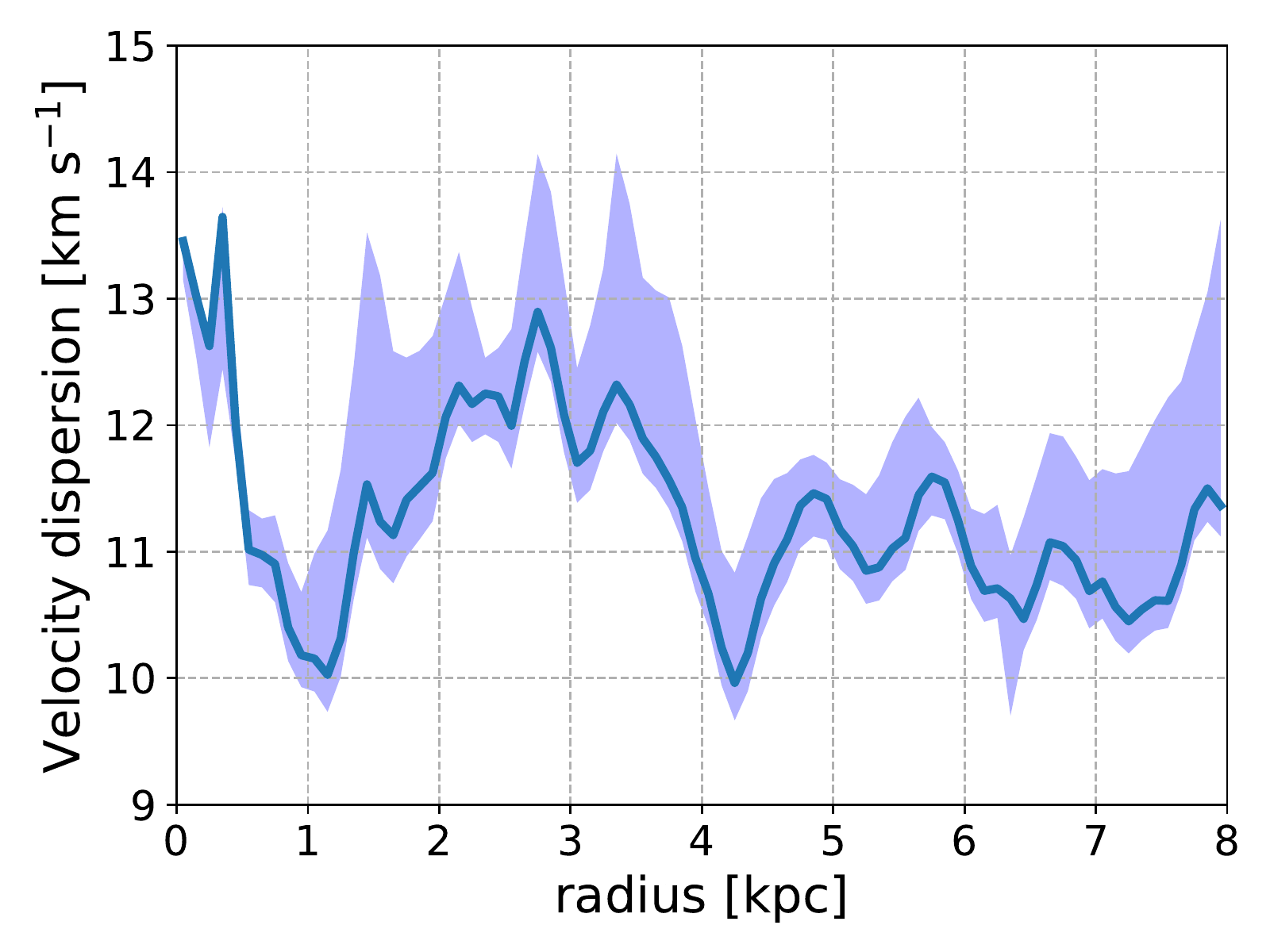}{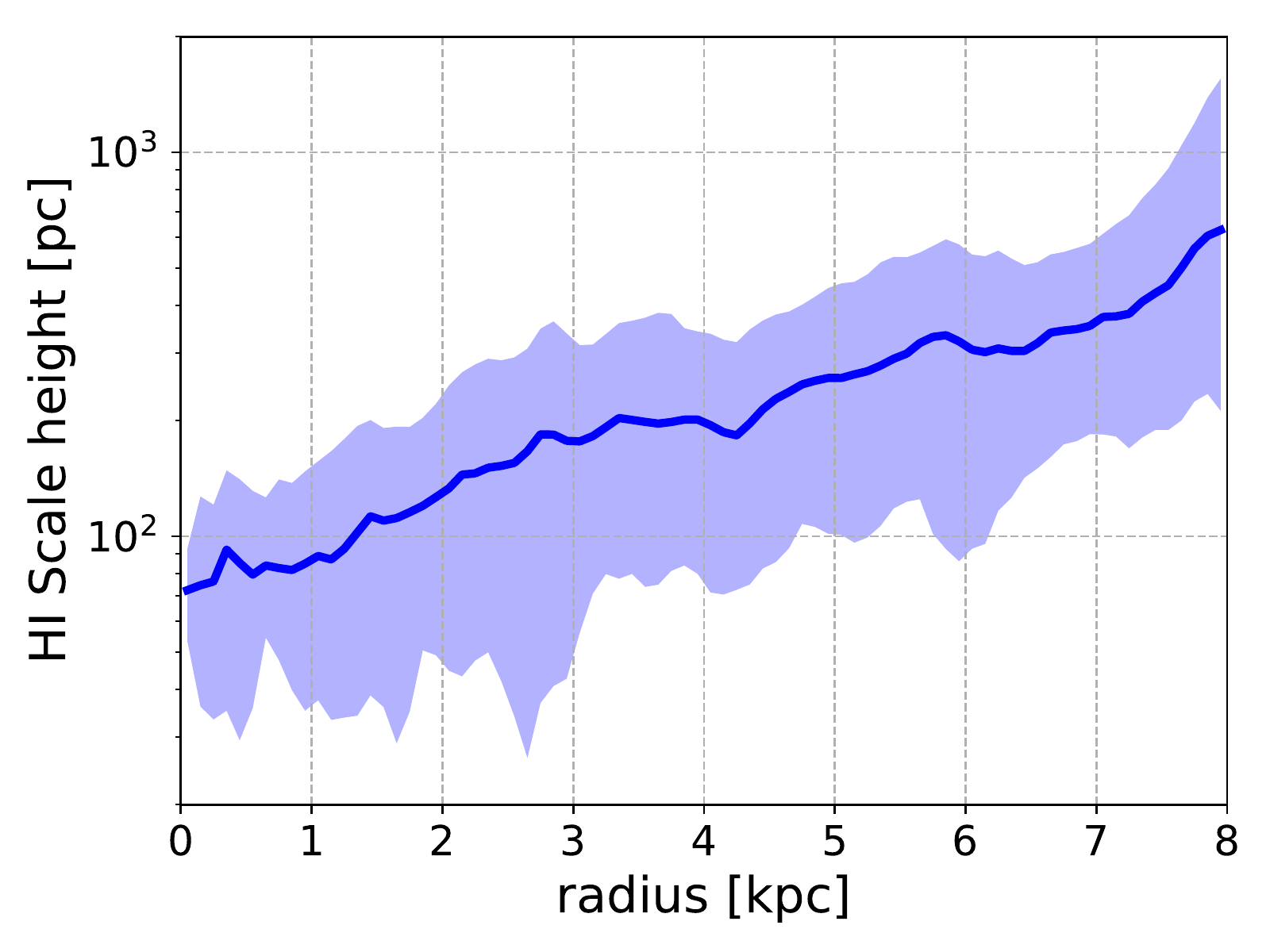}
\caption{Left: the \hi\ velocity dispersion (\hidisp) as a function of radius, calculated from the stacked spectra of \hi\ in 100 pc radial bins. Right: The \hi\ scale-height derived from Equation~\ref{eq:height}, assuming a vertical hydrostatic equilibrium.}
\label{fig:hi_disp}
\end{figure*}

\subsection{Molecular Gas} \label{data:mol}

As part of the M33 CO Large Program \citep{gratier10,druard14}, the CO(2--1) line has been observed over the whole disk of M33 down to a noise level of 20 mK per channel. The On-The-Fly mapping technique was done with the HERA multibeam dual-polarization receiver \citep{schuster04} on the IRAM 30-meter telescope on Pico Veleta, Spain. We adopt a line ratio CO(2--1)/CO(1--0) of 0.7 \citep[e.g.,][]{leroy08}. 

The data have a spatial resolution of $12''$ and a spectral resolution of 2.6 km s$^{-1}$. Since the CO map is used to identify individual GMCs, we do not convolve it to coarser resolution to match the \hi\ map. This practically has no effect because we determine the turbulent energy of atomic and molecular gas, separately.

\subsubsection{Mass Surface Density}

We multiply the CO luminosity with the CO(1--0)-to-H$_2$ conversion factor ($\alpha_{\rm CO}$) to derive the molecular gas mass. We consider two cases for $\alpha_{\rm CO}$; a constant Galactic $\alpha_{\rm CO}$ of $4.3 \ M_\odot \ [{\rm K \ km \ s}^{-1} {\rm pc}^2]^{-1}$, and $\alpha_{\rm CO}$ that depends on the gas-phase metallicities and the total mass surface densities \citep[][and references therein]{bolatto13} as described below. For simplicity, we show the radial profile of $\Sigma_{\rm mol}$ with a constant Galactic $\alpha_{\rm CO}$ in Figure~\ref{fig:rad_prof} and record it in Appendix~\ref{app:table}. But, we apply variable $\alpha_{\rm CO}$ when comparing the turbulent and SNe energy in molecular clouds.

\subsubsection{CO-to-H$_2$ Conversion Factor}

Gas in the lower metallicity environment has lower dust-to-gas ratio, and hence, requires a higher gas column density to shield the gas to protect CO from dissociation \citep{vandishoeck88,wolfire10,glover11}. Thus, a higher conversion factor is required because there is less CO per H$_2$ molecule. \citet{bolatto13} gives a prescription to estimate this correction to the conversion factor as
\begin{equation} \label{eq:metal_corr}
\alpha_{\rm CO} = 2.9 \ {\rm exp} \left( \frac{0.4}{Z' \ \Sigma_{\rm GMC}^{100}} \right) \ M_\odot \ ({\rm K \ km \ s}^{-1} \ {\rm pc}^2)^{-1},
\end{equation}
where $Z'$ is the gas-phase metallicity relative to the Solar value of $12+$log(O/H)$= 8.7$ \citep{prieto01}, and $\Sigma_{\rm GMC}^{100}$ is the surface density of GMCs in units of $100~M_\odot$~pc$^{-2}$.

What is the gas-phase metallicity in M33? Recent measurement by \citet{cipriano16}, based on the electron temperature, found $12+$log(O/H) $= 8.52 - 0.36 (r/R_{25})$. This means the conversion factor varies from 1.2 times higher than the Galactic $\alpha_{\rm CO}$ in the center of M33 to 2.7 times higher than the Galactic $\alpha_{\rm CO}$ at $R_{25} \approx 7.7$ kpc. Those values of conversion factor are in agreement with what were found by \citet{leroy11}, where they used dust emission as a tracer for H$_2$ surface density, but the gradient is steeper than that measured by \citet{rosolowsky08}, thus gives us a more extreme case.

In addition to metallicities, the total surface density can also affect $\alpha_{\rm CO}$. This is because molecular gas may encompass both gas and stellar gravitational potential, especially in the denser regions, such as galactic center. This additional contribution from stellar component broadens the CO velocity dispersion, giving a false impression that the molecular gas surface density (which is proportional to surface brightness and velocity dispersion) is higher than what it should be. In other words, the molecular gas is not self-virialized, but instead, in virial equilibrium with gas and stellar gravity. This change in $\alpha_{\rm CO}$ in the galactic center has been inferred by \citet{sandstrom13} in nearby galaxies. \citet{bolatto13} suggest that $\alpha_{\rm CO}$ is related to the total surface density {\it alone} as $\alpha_{\rm CO} \propto (\Sigma_{\rm tot}/100 \ M_{\odot} \ {\rm pc}^{-2})^{-0.5}$ for $\Sigma_{\rm tot} > 100 \ M_{\odot}$ pc$^{-2}$. This surface density occurs inside 3 kpc radius in M33 (Figure~\ref{fig:rad_prof}), therefore, this $\alpha_{\rm CO}$ variation only affects clouds in the inner disk.

\subsubsection{Velocity Dispersion}

The CO velocity dispersion is calculated using the 2nd-moment method (the intensity-weighted square of the velocity) in {\tt MIRIAD} package \citep{sault95}. We blank channels with CO intensity less than 0.1 K (5 times the typical noise in the data cube), and only include $\pm 3$ channels (equivalent to 15.6 km s$^{-1}$) from the mean velocity. The value of this "window channel" is set to be larger than the typical velocity dispersion in the ISM ($\sim 10$ km s$^{-1}$), but also, not too wide to be contaminated by noise. The molecular gas kinetic energy is calculated using a formula analogous to Equation~\ref{eq:turb_energy}.

\subsection{Stellar Masses} \label{data:star}

The $K-$band radial profile is acquired from \citet{munoz-mateos07}, who used the $K-$band image of 2MASS Large Galaxy Atlas \citep{jarrett03}. To convert it to stellar mass surface density, we adopt a mass-to-light ratio ($M/L$) of $0.5 \ M_{\odot} \ L_{\odot, K}^{-1}$. The original image ($1''$ pixel size and $3''$ resolution) is resampled and convolved to match the \hi\ map with $4''$ pixel size and $12''$ resolution using the {\tt MIRIAD} package \citep{sault95}. The major uncertainty is the $M/L$, that has a factor of 2 variation in the K-band \citep{bell01}.

The profile is well fitted by a de Vaucouleurs profile for the inner 1 kpc and the exponential profile between 1 and 4 kpc. Therefore, we extend the stellar profile outside 5 kpc from the fit to the exponential disk profile. The map of \sigstar\ is shown in Figures~\ref{fig:gallery}. The radial profile of \sigstar\ is shown in Figure~\ref{fig:rad_prof} and is recorded in Appendix~\ref{app:table}.

\subsection{Star Formation Rates}

We retrieve the far ultraviolet (FUV) map from {\it GALEX} at effective wavelength of 1516 \AA \ \citep{depaz07} as a tracer of unobscured star formation energy, and the mid infrared map (MIR) from {\it Spitzer} MIPS 24$\mu$m \citep{rieke04,gordon05,dale09} as a tracer of obscured star formation energy that is reradiated by the dust. Both FUV and MIR maps have been background subtracted. We correct the FUV map for Galactic extinction of $E(B-V) = 0.0418$ \citep{schlegel98} and using a conversion of $A_{\rm FUV} = 7.9 \ E(B-V)$ \citep{depaz07}. The original resolutions of FUV and MIR maps are $4''.5$ and $6''$, respectively. Therefore, we convolve and regrid the FUV and MIR maps to match the \hi\ and CO maps using the {\tt MIRIAD} package \citep{sault95}.

The FUV and MIR surface brighness ($I_{\rm FUV}$ and $I_{\rm MIR}$) are converted to the star formation rate surface density ($\Sigma_{\rm SFR}$) using \citep{leroy08}
\begin{equation} \label{eq:sfr_conv}
\Sigma_{\rm SFR} = (8.1 \times 10^{-2} \ I_{\rm FUV} \ + \ 3.2 \times 10^{-3} \ I_{\rm MIR}) \ {\rm cos}(i),
\end{equation}
where $\Sigma_{\rm SFR}$ is in units of $M_{\odot}$ pc$^{-2}$, and both $I_{\rm FUV}$ and $I_{\rm MIR}$ are in units of MJy sr$^{-1}$. Equation (\ref{eq:sfr_conv}) assumes a \citet{kroupa01} Initial Mass Function (IMF), which is a factor of 1.59 lower than a \citet{salpeter55} IMF. The map of \sigsfr\ are shown in Figures~\ref{fig:gallery}. Its radial profile is shown in Figure~\ref{fig:rad_prof} and is recorded in Appendix~\ref{app:table}.

\section{Results} \label{sec:result}

\subsection{Separating Thermal and Turbulent Energies}
\label{sec:separate}

The total kinetic energy of the gas consists of thermal plus turbulent energy. In order to calculate the thermal energy density ($e_{\rm th} \propto n T$), we need to know the volume density ($n$) and temperature ($T$) of the gas. For our study, it is necessary to consider two-phases ISM: cold and warm neutral media \citep[CNM and WNM;][]{field65}, because their density and temperature can vary by about two orders-of-magnitude. Hence, the mass fraction of \hi\ in each of these media affect the resulting thermal energy.

Calculating the physical state of the gas ($n$ and $T$) requires calculating its thermal and chemical equilibrium states, which is beyond the scope of this paper. Therefore, we adopt the result from \citet{wolfire03}, where they calculated $n$ and $T$ for CNM and WNM as a function of galactocentric radius in the Milky Way (MW). We therefore assume that M33 is a miniature version of the MW. In particular, the thresholds of $n$ and $T$ for CNM and WNM at $0.5 \ R_{25}$ in the MW is assumed to be the same as those at $0.5 \ R_{25}$ in M33. We adopt $R_{25} = 16$ kpc for the MW \citep{bigiel12} and $R_{25} = 7.7$ kpc for M33 \citep{gratier10}.

In practice, the gas thermal energy is determined through the following steps. First, we calculate the \hi\ number density for each pixel in M33 as $n_{\rm \hi} = \Sigma_{\rm \hi} \ (2 h_{\rm \hi})^{-1}$, where $h_{\rm \hi}$ is the scale-height of the \hi\ gas (derived using Equation~\ref{eq:height}). Then, we compare $n_{\rm \hi}$ with the mean values of $n_{\rm CNM}$ and $n_{\rm WNM}$ in the MW as calculated by \citet{wolfire03} at the same radius in $R_{25}$. There are three possible outcomes of this comparison. (1) For $n > n_{\rm CNM}$, we assume all of the \hi\ mass is in CNM. (2) If $n < n_{\rm WNM}$, then most of the volume is in WNM. For simplicity, we assume that all the \hi\ mass is in WNM. (3) For $n_{\rm CNM}$ $\leq n \leq n_{\rm WNM}$, we assume both phases exist, and distribute the gas mass to be half CNM and half WNM \citep[as observed in the MW by][]{heiles03}.

\begin{figure*}
\centering
\epsscale{1.15}
\plottwo{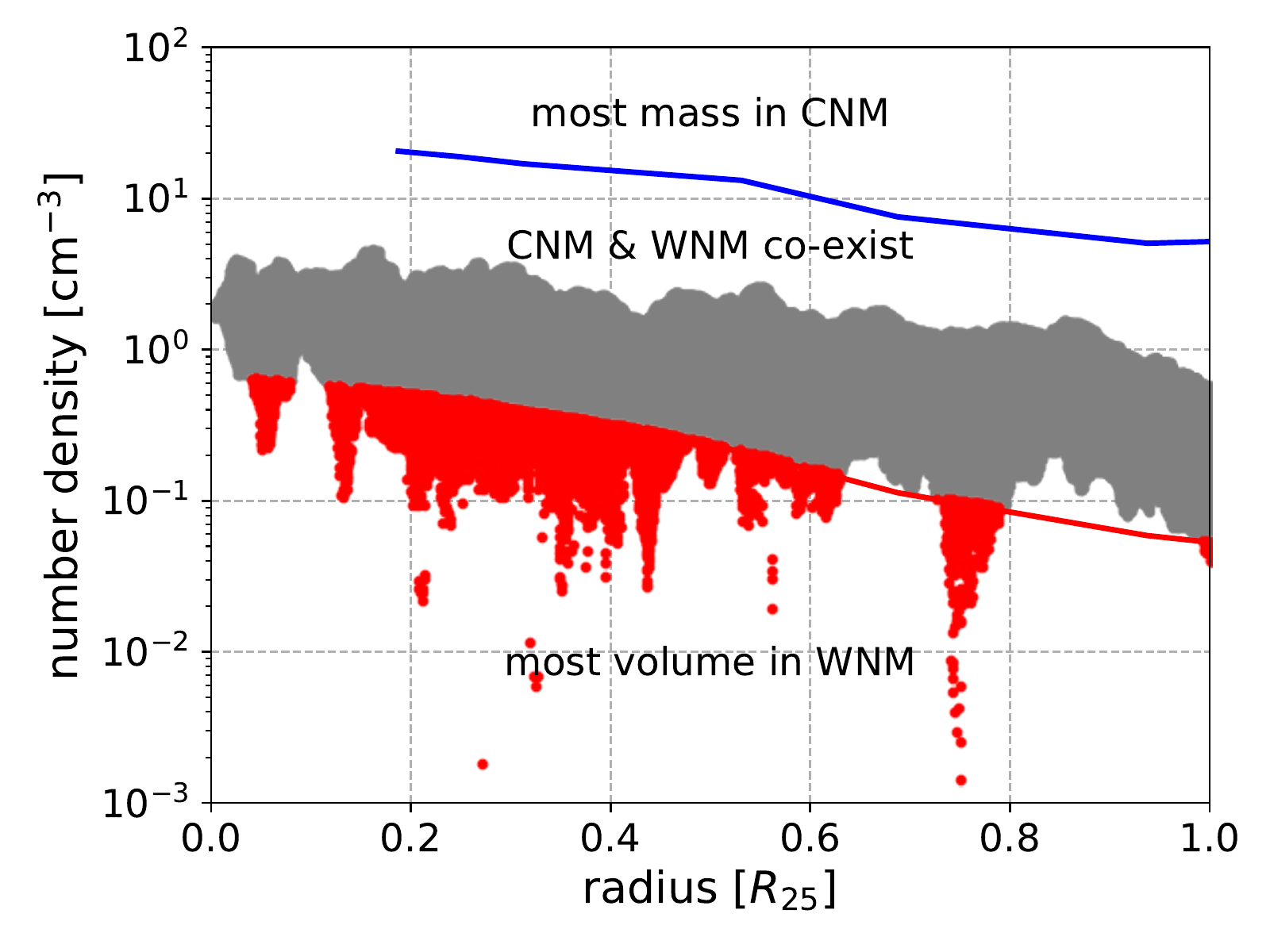}{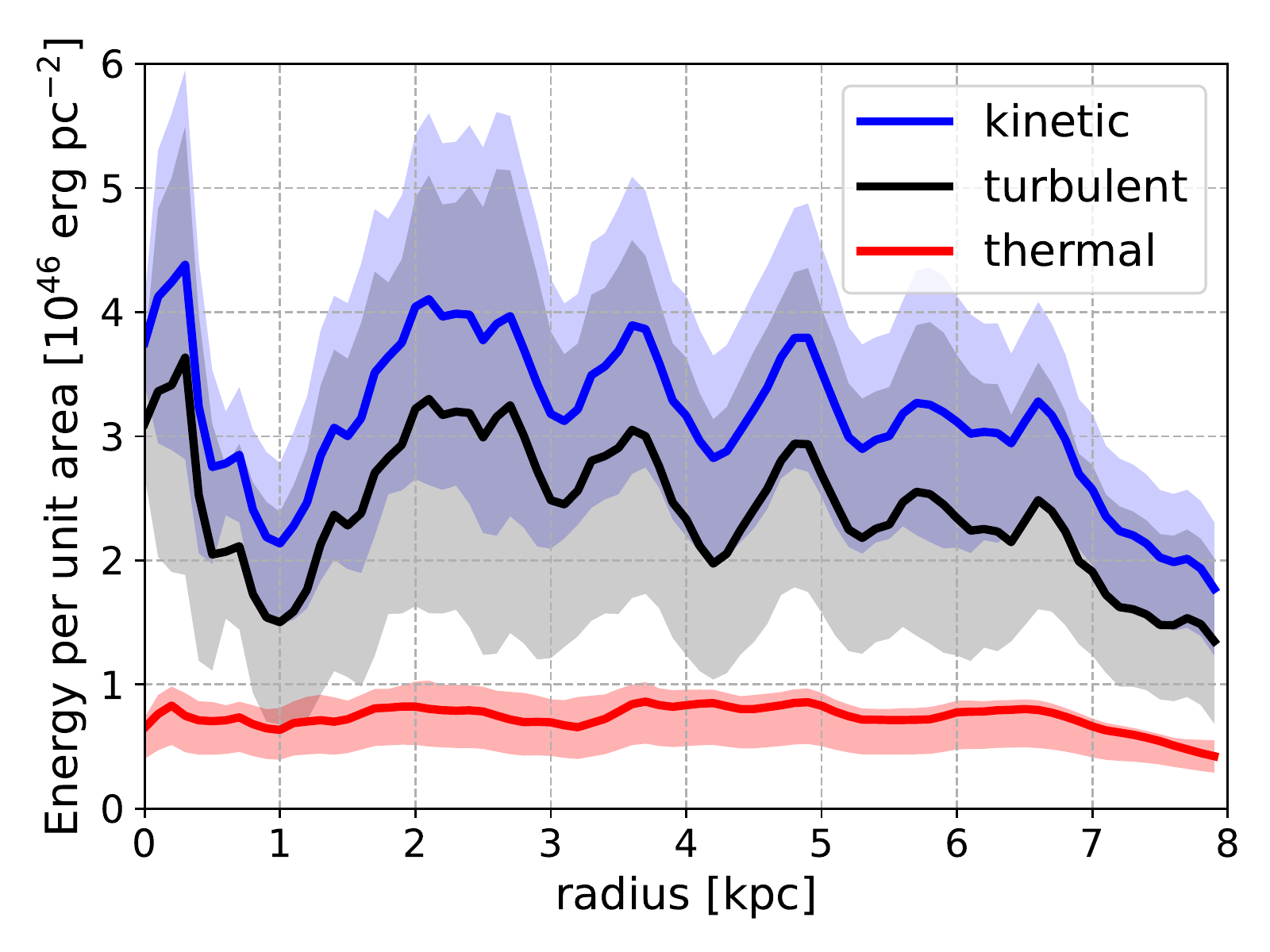}
\caption{Left: the number density of \hi\ in M33. The blue and red lines are the predicted number density of CNM and WNM, respectively \citep{wolfire03}. There are no data for CNM and WNM inside $0.2~R_{25}$, therefore, we extrapolate the red and blue lines using Univariate Spline in {\tt Scipy}. Both CNM and WNM co-exist at all radii (gray points), consistent with that in the Milky Way \citep{heiles03}. We assign \hi\ with number density below the red line as WNM (red points). Right: the radial profile of kinetic energy (blue line), turbulent energy (black line), and thermal energy (red line). The turbulent energy dominates over thermal energy at all radii by a factor of $\sim~2-3$. The uncertainties (shared regions) are derived from the error propagations.}
\label{fig:thermal}
\end{figure*}

In the left panel of Figure~\ref{fig:thermal}, we show the volume densities of \hi\ as a function of radius. The blue line marks the expected volume densities of CNM in the MW, while the expected number densities of WNM in the MW is marked as the red line. For most of the \hi\ mass, the number density of \hi\ in M33 are in the intermediate density between $n_{\rm WNM}$ and $n_{\rm CNM}$.

We add the thermal energy from the CNM and WNM as the gas thermal energy via
\begin{equation}
e_{\rm th} = \frac{3}{2} k_{\rm B} (n_{\rm CNM} T_{\rm CNM} + n_{\rm WNM} T_{\rm WNM}).
\end{equation}
Then, we subtract the thermal energy from the kinetic energy to get the turbulent energy. These thermal, kinetic, and turbulent energies are shown as the red, blue, and black lines, respectively, in the right panel of Figure~\ref{fig:thermal}. Their values are recorded in Appendix~\ref{app:table}. We see that the turbulent energy dominates over the thermal energy by a factor of $\sim 2-3$ at all radii. This means the driving mechanism of turbulence (e.g. stellar feedback or MRI) is still needed, even at the outermost radius where the star formation rate is negligible.

\subsection{Turbulence in Atomic Gas} \label{sec:atom_turb}

\begin{figure*}
\epsscale{1.15}
\plottwo{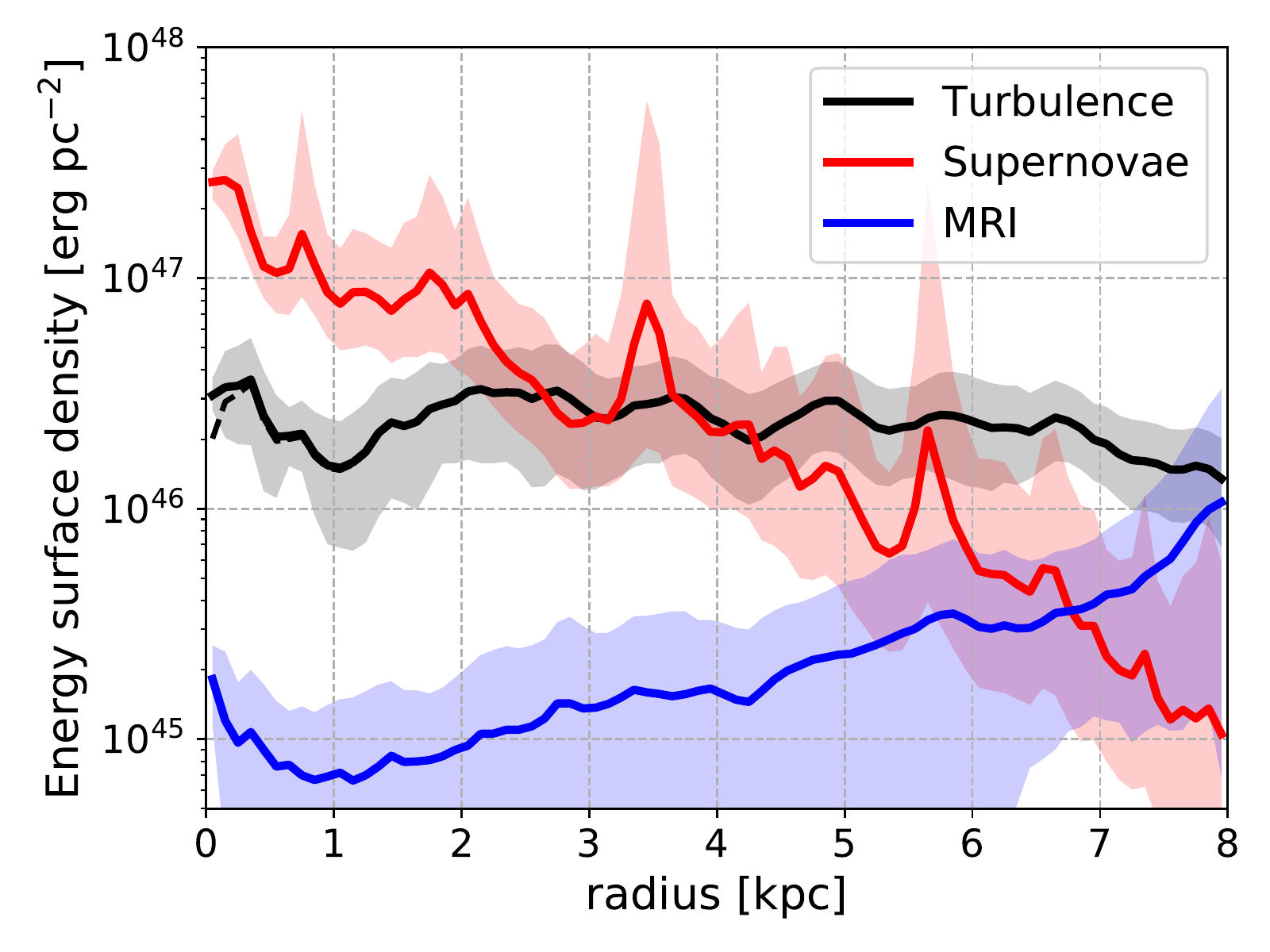}{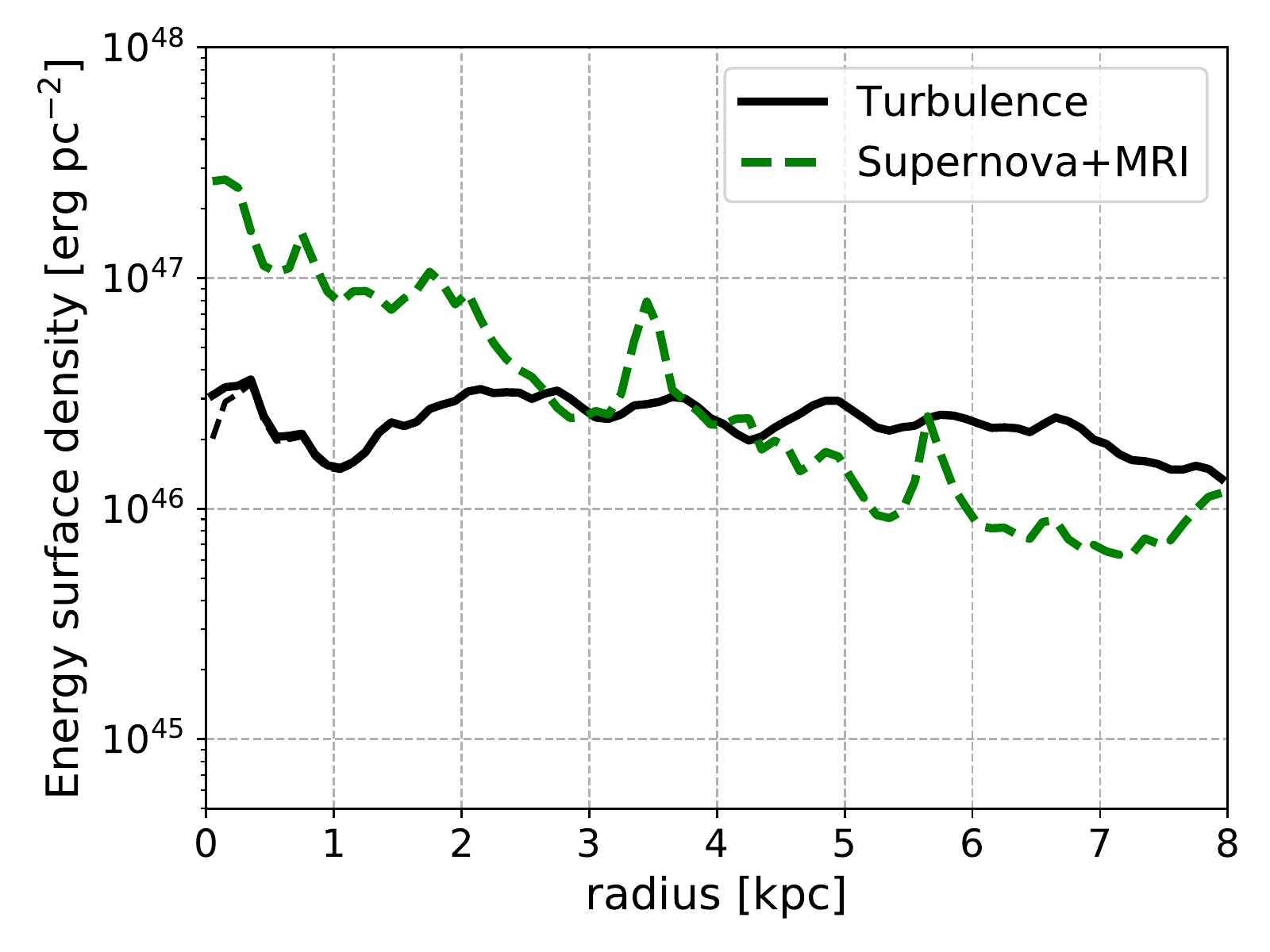}
\caption{Left: the radial profiles of turbulent energy (black line), SNe energy (red line), and MRI energy (blue line) as a function of radius. The black dashed line in the center is the turbulent energy after correction from beam smearing (Appendix~\ref{app:smear}). Right: a comparison between the turbulent energy (solid line) and the SNe$+$MRI energy (dashed line). The SNe energy assumes a constant dissipation time of 4.3 Myr (Appendix~\ref{app:sne}). While SNe have enough energy to maintain turbulence inside $\sim 4$~kpc, other sources of energy are required to maintain turbulence in the outer parts of M33.}
\label{fig:comp_energy}
\end{figure*}

In the left panel of Figure~\ref{fig:comp_energy}, we compare the radial profiles of turbulent energy per unit area with the possible sources of turbulent energy (SNe and MRI) assuming 100\% coupling efficiency (i.e. all SNe and MRI energies are converted to turbulence). We tabulate these turbulent, thermal, SNe, and MRI energies in Appendix~\ref{app:table}. There are three key points of our findings as described below.

First, the turbulent energy (black line) is almost flat ($\sim 2-3 \times 10^{46}$ erg pc$^{-2}$) inside 7 kpc radius. This is due to the fact that both $\Sigma_{\rm \hi}$ and $\sigma_{\rm HI}$ are almost constant within 7 kpc. Beyond 7 kpc, both \hi\ mass surface density and \hi\ velocity dispersion decrease, so that kinetic and turbulent energies also drop to $\sim~1~\times~10^{46}$ erg pc$^{-2}$. In the center, there is a small drop (black dashed line) if we took into account the effect of beam smearing to broaden the velocity dispersion (Appendix~\ref{app:smear}).

Second, the SNe energy (red line) dominates over the MRI energy (blue line) inside 6.5 kpc radius. The radial decline of SNe energy is due to the radial decline of the SFR surface density (as shown in Figure~\ref{fig:rad_prof}). Conversely, MRI energy rises outwards, driven by the increase of \hi\ scale-height as a function of radius (as shown in Figure~\ref{fig:hi_disp}).

Third, {\it individually}, the SNe energy alone is able to maintain turbulence inside $\approx 4$ kpc (with 100\% of coupling efficiency), while the MRI alone does not have enough energy to maintain turbulence  inside 8 kpc radius. Therefore, the source of turbulence in the outer parts must be from other sources, because the sum of SNe and MRI energies (dashed green line in the right panel of Figure~\ref{fig:comp_energy}) is smaller than the turbulent energy. In $\S$\ref{sec:acc} and $\S$\ref{sec:grav}, we argue  that the kinetic energy from the accreted material is enough to maintain turbulence at outer radius. It is interesting to note that the values of MRI and turbulent energies are converging at 8~kpc, so we still can not rule out the importance of the MRI as a source of turbulence outside 8~kpc.

All calculations above assume a constant dissipation time of $\approx 4.3$ Myr for SNe energy (Appendix~\ref{app:sne}). If somehow, SNe energy is able to escape the \hii\ regions and tap its energy to diffuse atomic ISM, then the appropriate driving-scale may be the thickness of \hi\ gas, $h_{\rm \hi}$ (as in the case of MRI energy). Therefore, we also do analogous calculations, but this time with $\tau_D~\propto~h_{\rm \hi}~\sigma_{\rm \hi}$ \citep{maclow04}. Since $h_{\rm \hi}$ is larger than the size of \hii\ regions (by a factor of $\sim 2-15$) while $\sigma_{\rm \hi}$ is only $20-30\%$ higher than a fiducial value of 10~km~s$^{-1}$, then the dissipation time becomes longer, and the SNe energy, required to maintain turbulence, also increases.

\begin{figure*}
\epsscale{1.15}
\plottwo{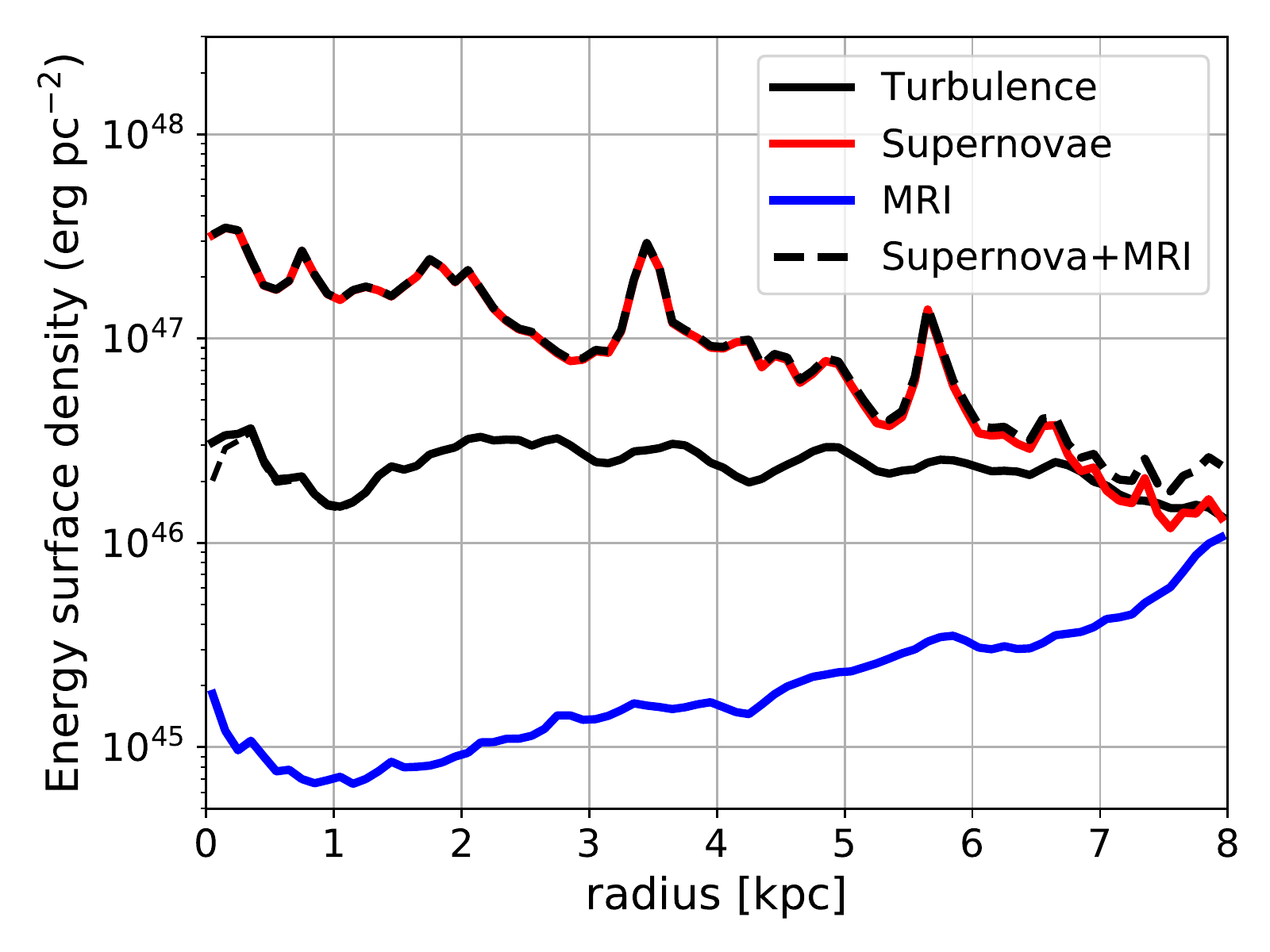}{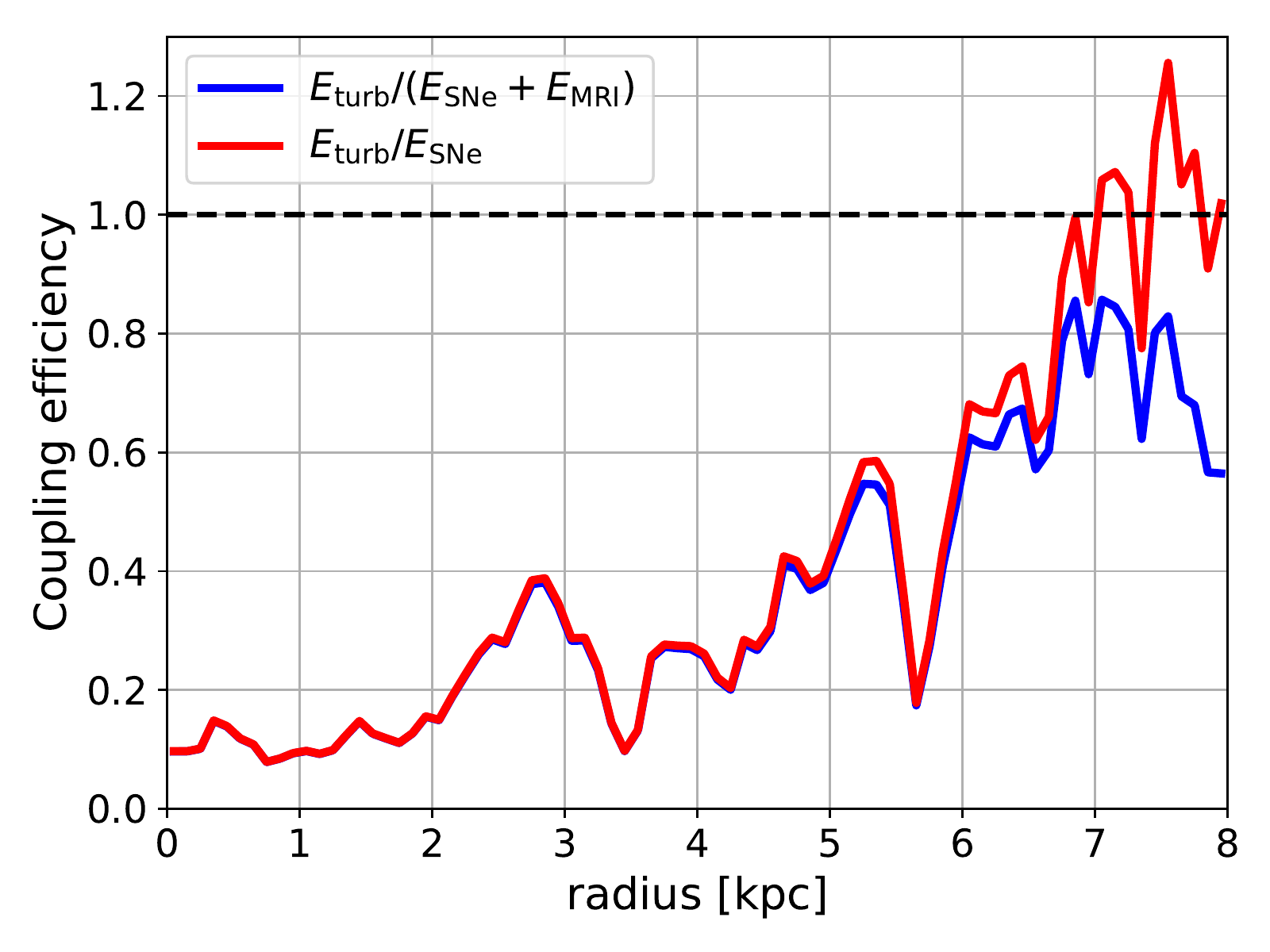}
\caption{Left: comparisons of energy surface density, similar to Figure~\ref{fig:comp_energy}, but for varying dissipation time $(\tau_D \propto h_{\rm \hi} \ \sigma_{\rm \hi}^{-1})$ in calculating the SNe energy. A larger driving-scale would result in a longer dissipation time and higher SNe energy. For clarity, we exclude the uncertainties in the Figure but they should be similar as in Figure~\ref{fig:comp_energy}. The black dashed line in the center is the turbulent energy after correction from beam smearing (Appendix~\ref{app:smear}). Right: coupling efficiency for the SNe and SNe+MRI energies. In this case, the combination of SNe and MRI energies is enough to feed turbulence at all radii, with coupling efficiency increases from $\sim 10\%$ in the center to $\sim 80\%$ in the outer part.}
\label{fig:comp_var}
\end{figure*}

In the left panel of Figure~\ref{fig:comp_var}, we compare this SNe energy (with $\tau_D~\propto~h_{\rm \hi}~\sigma_{\rm \hi}$), MRI energy, and turbulent energy. Unlike previous calculation with a constant $\tau_D$, now SNe have enough energy to maintain turbulence inside 7 kpc radius. Outside 7 kpc, the combination of SNe and MRI energies is required to be able to maintain turbulence. In this case, other sources of energy are not required to maintain turbulence. We define coupling efficiency as the ratio between the turbulent energy and the driving energy, i.e. the fraction of driving energy required to maintain turbulence. We show their values in the right panel of Figure~\ref{fig:comp_var}, where the coupling efficiency increases outward from $\sim 10\%$ to $\sim 80\%$.

A possible reason for the variation of coupling efficiency is the leakage of SNe energy through the SNe bubbles that flow out from the galaxy midplane to the intergalactic medium. This gas outflow, driven by stellar feedback, is usually strong enough to be observed in starburst galaxies \citep{bolatto13b,leroy15,martini18}. The \hi\ mass surface density is known to have small variation, while \sigsfr\ declines as a function of radius. This means SNe bubbles tend to overlap with each other near the galaxy center, which increases the likelihood of energy leakage from the galaxy. This process transfers SNe energy to the ionized medium out of the midplane, instead of being deposited into the \hi\ gas. In this view, the \hi\ gas only captures a fraction of SNe energy, and hence, leads to a smaller coupling efficiency in the center. Note one must take care in calculating the coupling efficiency. For example, if the magnetic field varies with radius, then the coupling efficiency of the MRI would have to vary as well, and if the magnetic field has a different value than we have assumed, then the average value of the coefficient would also differ from what we estimated here.

\subsection{Turbulence in Molecular Clouds} \label{sec:mol}

Since star formation occurs in molecular clouds, we also investigate whether the stellar feedback alone can maintain turbulence in molecular clouds. For this purpose, individual molecular clouds in M33 are identified through the following procedures. First, we utilize the {\tt CPROPS} package of \citet{rosolowsky06} to identify contiguous pixels in the CO data cube. These regions must have at least one pixel with SNR~$\geq 5$ and are bounded by pixels with SNR of 2 as their edges. The aim of this process is to separate signal from noise. As a result, we have a masked cube with binary values, zero for noise and one for signal.

We then collapse that masked cube along the velocity axis. Line-of-sights that only cover less than 3 channels are blanked because they are not sufficient for the calculation of velocity dispersion. Then, we label each contiguous region in this 2-dimensional map as an individual molecular cloud. We also remove clouds with the total number of pixels less than 15 (equivalents to an effective radius of 9.1 pc) because smaller clouds are susceptible to noise. At the end, we identify 124 molecular clouds in M33. This is fewer than 148 clouds that were cataloged by \citet{engargiola03} because our selection is more conservative and the \citeauthor{engargiola03} catalog also consists of many smaller clouds.

The kinetic and SNe energies within a molecular cloud are calculated by adding the respective energy from each pixel within the boundary of that molecular cloud, set by the masking process described before. We do bootstrap resampling to estimate their uncertainties. For a temperature of 10 K, the sound speed in molecular clouds is $\sim 0.2$ km s$^{-1}$, while our measured velocity dispersion is a few km s$^{-1}$ (Appendix~\ref{app:table}). Therefore, $\sigma_{\rm turb} \gg \sigma_{\rm therm}$, and hence, we can approximate $E_{\rm turb} \approx E_{\rm kin}$. However, keep in mind that the measured SNe energy within the molecular cloud is probably an overestimate because the stars will move away from the clouds in a time-scale of Myrs for those stars to evolve to the end of their lives and the ionizing radiation from the stars pushes the gas away from the stars \citep{mckee84}.

The comparison between $E_{\rm turb}$ and $E_{\rm SNe}$ for each molecular cloud is shown in Figure~\ref{fig:cloud_turb} and tabulated in Appendix~\ref{app:table}. The molecular turbulent energy per cloud (blue points) is correlated with the supernovae energy per cloud, with $\approx 0.83\%$ of median coupling efficiency, defined as the ratio of $E_{\rm turb}$ over $E_{\rm SNe}$. This value is only slightly different when $\alpha_{\rm CO}$ that depends on metallicity and mass surface density is adopted (0.92\%). Therefore, we conclude that supernovae have enough energy to maintain turbulence in molecular clouds.

\begin{figure}
\centering
\includegraphics[width=0.45\textwidth]{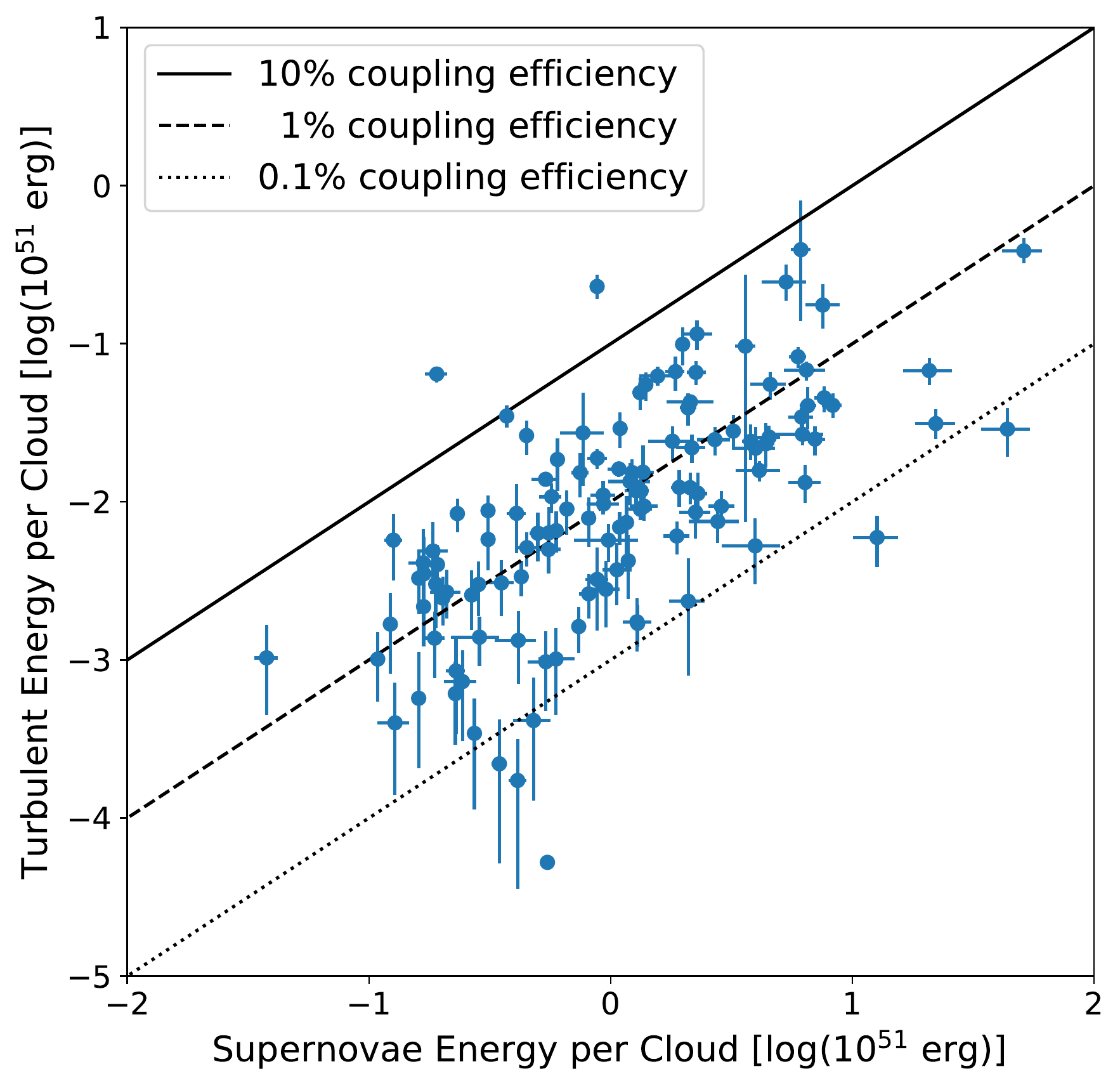}
\caption{A comparison between the turbulent energy and SNe energy in the molecular clouds. The solid, dashed, and dotted black lines mark the SNe coupling efficiency of 10\%, 1\%, and 0.1\%, respectively. This figure shows that SNe energy is able to maintain turbulence in the molecular clouds.}
\label{fig:cloud_turb}
\end{figure}

Note that the kinetic energy per cloud in Figure~\ref{fig:cloud_turb} is roughly in agreement with the simulations outcome from \citet{padoan16}, where their total kinetic energy (integrating over the whole simulation volume) is $\sim 10^{50}$~erg for ISM above a density of 100~cm$^{-3}$. Within their simulation volume, there are 10 clouds with masses $\gtrsim 10^4~M_{\odot}$ (typical of GMCs), so that their kinetic energy per cloud is $\sim 10^{49}$~erg.

As a check, we also loosen  the constraint for cloud identifications by reducing the peak SNR to be 2, while keeping the SNR at the edges as is. Also, there is no decomposition for the contiguous regions. In other words, we are likely to identify Giant Molecular Associations (GMAs) rather than GMCs. The aim is to include a more diffuse CO emission with low star formation rate, so that we can check whether the SNe energy can still maintain turbulence per GMA. As in a GMC, the SNe energy per GMA is derived by adding SNe energy from all pixels within the boundary of a GMA. We find that the correlation between molecular turbulent energy and SNe energy still exists, with a median coupling efficiency of $\approx 0.64\%$ for a Galactic $\alpha_{\rm CO}$ and 0.73\% for variable $\alpha_{\rm CO}$. The fact that it is still less than $100\%$ means the SNe energy can still maintain turbulence, even after the inclusion of more diffuse CO emission.

\section{Discussion} \label{sec:discuss}

\subsection{Comparisons with Previous Works}

We show that SNe have enough energy to maintain turbulence inside radius of $\approx 4$ kpc in M33 (for a constant dissipation time of 4.3 Myrs). This finding is, in general, consistent with the previous study by \citet{tamburro09} in a sample of 11 nearby disk galaxies selected from the THINGS survey \citep{walter08}. However, there are at least four differences between \citeauthor{tamburro09} and our results as listed below.

\begin{enumerate}

\item Their measured $\sigma_{\rm HI}$ is rising towards the center and could reach a value of $\sim 25$ km s$^{-1}$. This could be the effect of beam smearing because their physical resolution is $\sim 500$ pc at 10 Mpc distance (typical distance of their targets). Our resolution is one order-of-magnitude better ($\approx 80$ pc) than that in \citeauthor{tamburro09}, therefore, we do not find an increase of $\sigma_{\rm HI}$ towards the center.

\item Their measurement covered at least twice $R_{25}$, while we only cover up to $R_{25}$. Unlike our conclusion that SNe energy is larger than turbulent energy inside $\sim~0.5~R_{25}$, \citeauthor{tamburro09} found that it occurs inside $R_{25}$. Eventhough \citeauthor{tamburro09} adopt a dissipation time of 9.8 Myr (a factor of two longer than our fiducial value), this difference still can not explain the discrepancy between SNe and turbulent energies outside $0.5~R_{25}$. Only when we calculate dissipation time as $\tau_{\rm D} = h_{\rm \hi} \sigma_{\rm \hi}^{-1}$, the SNe have enough energy to maintain turbulence within $\sim R_{25}$.

\item \citeauthor{tamburro09} mentioned that MRI has enough energy to maintain turbulence outside $R_{25}$. The $R_{25}$ value in M33 is 7.7 kpc \citep[from Hyperleda database;][]{makarov14}, and our data are restricted inside 8 kpc because of our sensitivity limit. Also beyond 8 kpc, \hi\ is strongly warped, making the measurements of \hi\ velocity dispersion and rotation curve become difficult. Hence, we cannot compare directly to \citeauthor{tamburro09} result. Instead, we note that MRI energy is less than turbulent energy inside $R_{25}$, but interestingly, both energies are converging at 8 kpc (see left panel of Figure~\ref{fig:comp_energy}).

\item Galaxies in the \citeauthor{tamburro09} sample have rotation speed around $200-300$ km s$^{-1}$ in the flat part \citep{blok08}, while the peak of rotation velocity in M33 is $110$ km s$^{-1}$ (Appendix~\ref{app:kin}). This means, for a given radius in the flat part, $\Omega$ and shear rate are higher in their sample. Since MRI energy is proportional to the shear rate (Equation~\ref{eq:mri_final}), this gives rise to higher MRI energy in their sample. On the other hand, \hi\ mass surface density and velocity dispersion are comparable between M33 and their disk galaxies, which means their \hi\ turbulent energies are also comparable with ours. Altogether, these differences may explain why the MRI has enough energy to maintain turbulence in the disk galaxies, but not in dwarfs, such as M33.
\end{enumerate}

Another extensive study about turbulence in galaxies was conducted by \citet{stilp13} in a sample of dwarf galaxies. They found that the SFR is able to maintain turbulence in regions where $\Sigma_{\rm SFR} \gtrsim 0.1 \ M_\odot$ Gyr$^{-1}$ pc$^{-2}$, which is equivalent to $\Sigma_{\rm ESN} \gtrsim 2 \times 10^{45}$ erg pc$^{-2}$ for $\epsilon_{\rm SN} = 1$ (Equation~\ref{eq:sne_final}). On the other hand, $\Sigma_{\rm ESN} \gtrsim 2 \times 10^{46}$ erg pc$^{-2}$ is needed to maintain turbulence in M33 (see Figure~\ref{fig:comp_energy}), an order-of-magnitude higher than their threshold. \citeauthor{stilp13} also mentioned that the MRI energy is {\it unable} to maintain turbulence in regions of low star formation rates (consistent with our finding), because the velocity dispersion of \hi\ in dwarf galaxies is similar to the outer disk of spirals, but dwarf galaxies have less shear (and hence less MRI energy) compared to that in spiral disks.

\subsection{Tidal interaction} \label{sec:acc}

M33 and M31 are known to be interacting, with a `bridge` of \hi\ gas connecting those two galaxies is detected \citep[e.g.,][]{braun04,putman09,lockman12}. Does their tidal interaction generate enough energy to feed turbulence? The rate of energy injected by accretion can be estimated simply as the kinetic energy of the accreted materials \citep{klessen10}, $\dot E_{\rm acc}=0.5~\dot M_{\rm acc}~V_{\rm acc}^2$, where $\dot M_{\rm acc}$ and $V_{\rm acc}$ are the accreted mass rate and the accretion velocity. For a galaxy with size $R \sim 10$ kpc (comparable to that in M33) and turbulent dissipation time $\tau_D \sim 9.8$ Myrs, the energy surface density due to accretion is
\begin{equation}
\Sigma_{\rm acc} = 0.5~\dot M_{\rm acc}~V_{\rm acc}^2~\tau_{\rm D}~(\pi R^2)^{-1} \sim 1 \times 10^{46} \ {\rm erg \ pc}^{-2},
\end{equation}
where we adopt $\dot M_{\rm acc} = 3 \ M_\odot$ yr$^{-1}$ and $V_{\rm acc} = 100$ km s$^{-1}$ for M33 as reported by \citet{zheng17}. This energy density is comparable to the turbulent energy density at the outer radius. Therefore, tidal interaction is a possible source of turbulent energy in the outer part of M33. However, note that those values of $\dot M_{\rm acc}$ and $V_{\rm acc}$ are measured outside our radial range ($>8$~kpc).

\begin{figure}
\centering
\includegraphics[width=0.5\textwidth]{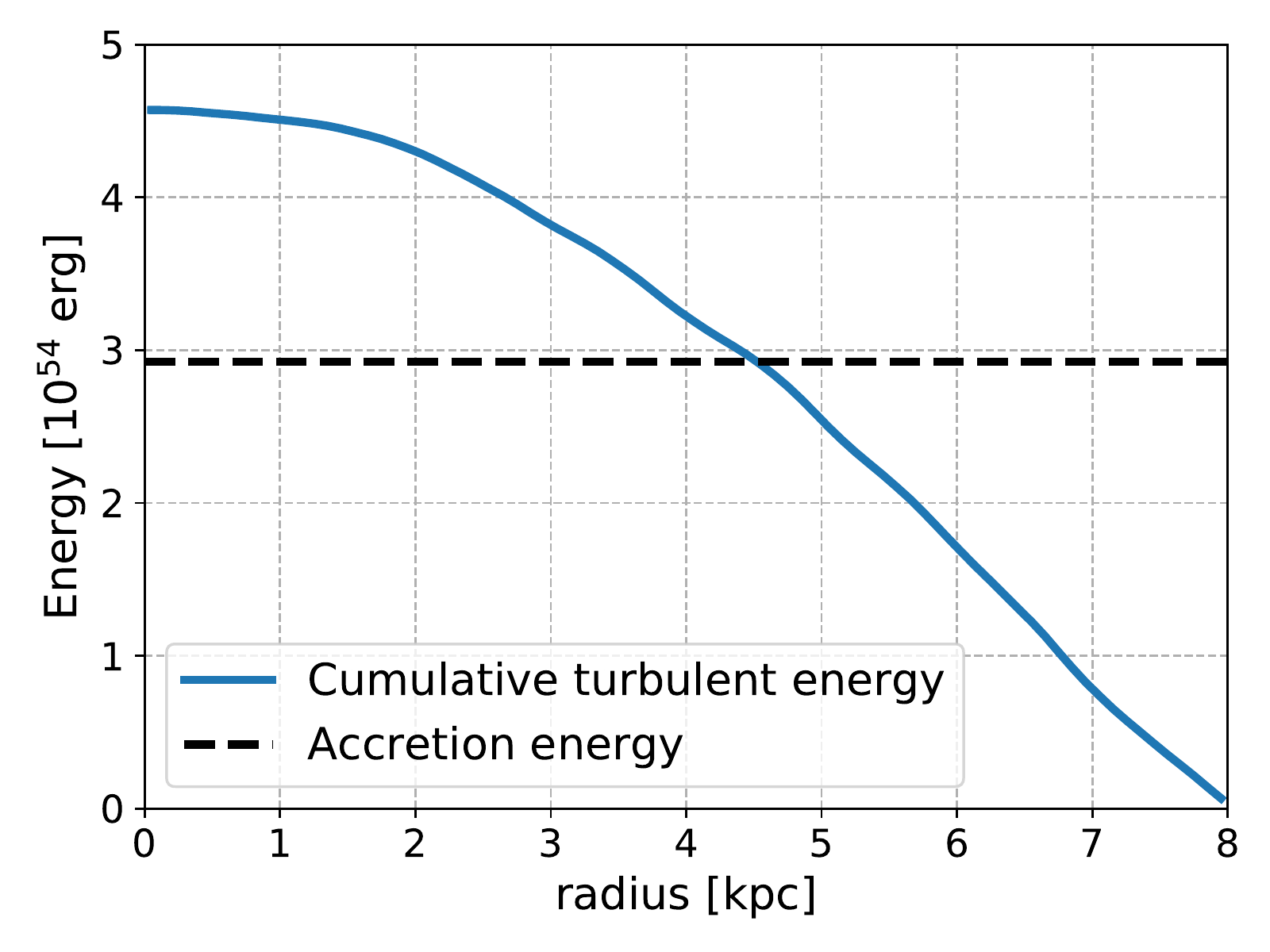}
\caption{A cumulative turbulent energy of atomic gas, {\it integrated from outside to inside} (blue line). As a comparison, the total kinetic energy from accretion in a period of 9.8 Myrs (dissipation time) is shown as a dashed black line. This means the accretion is able to maintain turbulence outside 5 kpc radius of M33, assuming a 100\% coupling efficiency.}
\label{fig:acc_energy}
\end{figure}

Figure~\ref{fig:acc_energy} shows the cumulative turbulent energy as the blue curve. This cumulative energy is calculated from {\it outside to inside} because the accretion kinetic energy is originated from outside the galaxy. As a comparison, we also show the total kinetic energy from accretion within a dissipation time-scale of 9.8 Myrs. If accretion is able to maintain turbulence with 100\% coupling efficiency, then accretion {\it alone} may be the source of turbulent energy {\it outside} $\approx 4.5$ kpc radius of M33. Inside 4.5 kpc, there is not enough energy from accretion to maintain turbulence. However, as we mentioned in $\S$\ref{sec:atom_turb}, the SNe energy is able to maintain turbulence in the inner region of M33.

\subsection{Gravity-driven Turbulence}
\label{sec:grav}

While the kinetic energy generated by the gas accretion is large enough to account for turbulent energy, we remain skeptical on how this energy can generate turbulence in the inner disk of M33, where the mass inflow rate is much smaller than the mass accretion rate \citep{wong04,schmidt16}. Also, if the accreting gas has a temperature much less than $10^6$ K, then the interaction of the accreted gas with gas in the disk will result in a radiative shock, and most of the energy will be radiated away, resulting in a very small value of coupling efficiency.

\citet{krumholz10} and \citet{krumholz16} proposed a scenario that the gravitational instability from the inflowing materials through the disk can generate turbulence. While this may be true for large gas velocity dispersion ($\sigma_{\rm gas} \gtrsim 50$ km s$^{-1}$) and small gas fraction as in active star-forming galaxies, the gas velocity dispersion in M33 is small ($\lesssim 10$ km s$^{-1}$), and thus, the separation between the stellar feedback and gravitational instability as the source of turbulence is indistinguishable (their Figures 1 and 2).

Since the actual mass accretion in the disk of M33 is unknown, we test this gravity-driven turbulence by using toy models of gas accretion rate, $\dot M$, as follows.
\begin{enumerate}
\item A constant $\dot M$ throughout the disk radii, with any values less than $3~M_\odot$~yr$^{-1}$. In other words, we treat the measured $\dot M$ by \citet{zheng17} as an upper limit. Physically, this can be interpreted as a zero net flux of gas in each radial annulus, except at the center of galaxy where the gas is accumulated.
\item A constant $\dot M$ per unit area. The annulus area in the outer part is larger than the inner part, therefore in this model, the accretion rate decreases inward, in general agreement with the results of \citet{schmidt16}. There are two branches of this model: (a) the total accretion rate must not exceed $3 \ M_\odot$ yr$^{-1}$ \citep{zheng17}, meaning that all the gas accretion is originated from outside the galaxy, and (b) the inner part of accretion is a scale-down version of the accretion in the outermost part of the galaxy (8 kpc), which is set to be $\dot M_{\rm out} = 3 \ M_\odot$ yr$^{-1}$. In other words, we use a scaling relation $\dot M_{\rm in}/\dot M_{\rm out} = A_{\rm in}/A_{\rm out}$, where $A$ is the annulus area.
\end{enumerate}
For each of those models, we can calculate the turbulent energy induced by gravity by using Equation~\ref{eq:grav}.

Another parameter in this model is the \citet{toomre64} $Q$, which can be calculated by two different ways: (a) set it to be unity, and (b) using the \citet{wang94} approximation, which is done by \citet{krumholz16}, for both stars and gas as
\begin{equation} \label{eq:q}
Q \approx \frac{\sqrt{2}}{\pi G} \frac{V_c \ \sigma \ f_g}{r \ \Sigma},
\end{equation}
where $V_c$ is the circular speed of the galaxy (derived in Appendix~\ref{app:kin}), $f_g$ is the total gas fraction, i.e. $\Sigma/(\Sigma + \Sigma_{\star})$, and $r$ is the galactocentric radius.

\begin{figure*}
\epsscale{1.15}
\plotone{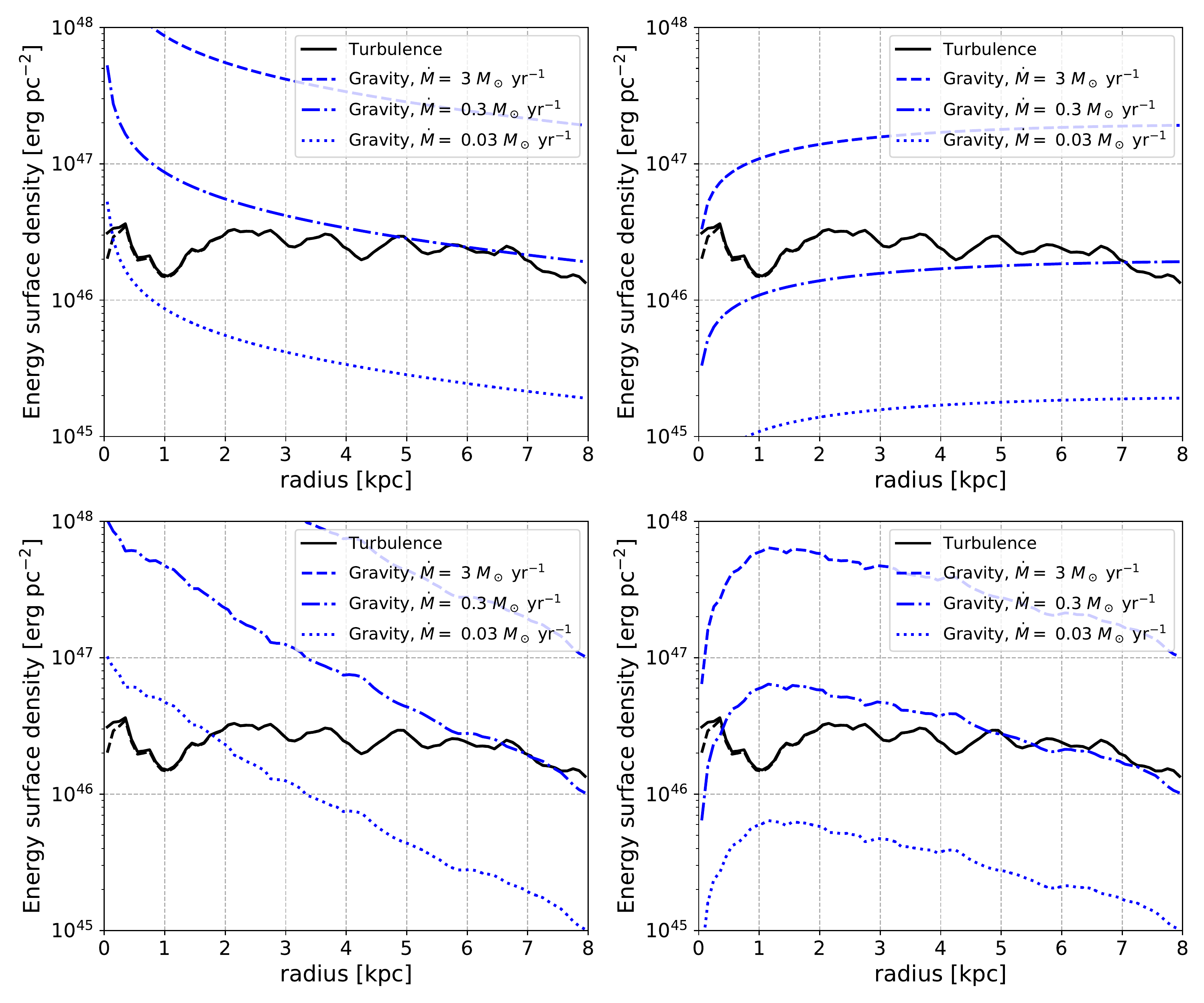}
\caption{Comparisons between the turbulent energy (black) and the gravito-turbulent energy (blue) generated by accretion rate, $\dot M$, for various models: a constant $\dot M$ and $Q = 1$ (top left), a constant $\dot M$ and variable $Q$, calculated using Equation~\ref{eq:q} (bottom left), a constant $\dot M$ {\it per unit area} and $Q = 1$ (top right), and a constant $\dot M$ {\it per unit area} and variable $Q$ (bottom right). The magnitude of gravito-turbulent energy depends on $\dot M$, therefore, we vary $\dot M$ to be 3, 0.3, and 0.03 $M_\odot$ yr$^{-1}$. In all models, $\dot M \gtrsim 0.3 \ M_\odot$ yr$^{-1}$ is required for the gravito-turbulent energy to maintain turbulence in M33.}
\label{fig:grav}
\end{figure*}

In Figure~\ref{fig:grav}, we compare the observations of turbulent energy in atomic gas (black lines) with the outcome of those models (blue lines). Each row has different assumption on the $Q$ parameter (top row for $Q = 1$ and bottom row for variable $Q$), while each column has different model of $\dot M$ (left column for a constant $\dot M$ and right column for a constant $\dot M$ per unit annulus area). The strength of the gravito-turbulent energy, $\Sigma_{\rm grav}$, depends on $\dot M$ (for constant $\dot M$ models) or $\dot M_{\rm out}$ (for models with constant $\dot M$ per unit area), so we vary that value to be 3, 0.3, and 0.03 $M_\odot$ yr$^{-1}$, shown as the dashed, dot-dashed, and dotted curves, respectively.

For all models, the values of $\Sigma_{\rm grav}$ with $\dot M = 3~M_\odot$~yr$^{-1}$ exceeds $\Sigma_{\rm turb}$ by about an order of magnitude, while $\Sigma_{\rm grav}$ with $\dot M=0.3~M_\odot$~yr$^{-1}$ has similar energy as $\Sigma_{\rm turb}$. Since we do not know the actual value of the coupling efficiency of $\Sigma_{\rm grav}$, we can only give a loose constraint that $\dot M\gtrsim~0.3~M_\odot$~yr$^{-1}$ is required for $\Sigma_{\rm grav}$ to be the sole driver of turbulent in M33.

The trend of $\Sigma_{\rm grav}$ as a function of galactocentric radius is also of particular interest. If $\Sigma_{\rm grav}$ is the sole driver of turbulence, then the trends should mimic that of $\Sigma_{\rm turb}$, i.e. relatively flat as a function of radius. All models, except for the constant $\dot M$ with variable $Q$ (bottom left in Figure~\ref{fig:grav}), show a relatively flat trend outside $\sim 1$ kpc radius. However, a detailed measurement of the inflowing mass as a function of radius inside the disk of M33 \citep[similar to the work of][]{schmidt16} is required to further constrain those models.

\section{Summary} \label{sec:summary}

M33 is a prospective place to study the interstellar turbulence, given the wealth of archival, high-resolution, multi-wavelength data. Here, we investigate the origin of turbulence in the diffuse \hi\ gas and in the molecular clouds, with respect to three possible sources; the stellar feedback from supernovae (SNe), the magneto-rotational instability (MRI) that is generated by the differential rotation of the galaxy, and the gravity-driven turbulence from accreted materials. 

The two-phase model of the ISM in the Milky Way \citep{wolfire03} is adopted to calculate the fraction of \hi\ gas mass in the WNM and CNM phases. The thermal energy is estimated as the sum of WNM and CNM thermal energies. Then, the turbulent energy is derived from the kinetic energy of \hi\ gas, after subtraction of its thermal energy. As a result, we find that the turbulent energy is a factor of $\sim 2-3$ times higher than the thermal energy at all radii (Figure~\ref{fig:thermal}).

By comparing the turbulent energy of atomic gas against the SNe and MRI energies in radial bins, we show that SNe have enough energy to maintain turbulence inside $\approx 4$ kpc, while the MRI does not have enough energy to maintain turbulence at inside 8~kpc (Figure~\ref{fig:comp_energy}). Therefore, another source of energy is required at the outer parts on M33. However, when we allow the turbulent dissipation time to vary according to the scale-height and velocity dispersion of \hi\ gas, SNe energy is able to maintain turbulence out to 7 kpc, while the sum of SNe and MRI energies can sustain turbulent inside 8~kpc (Figure~\ref{fig:comp_var}). For later case, the fraction of SNe$+$MRI energy that is needed to maintain turbulence, i.e. their coupling efficiency, rises from $\sim 10\%$ in the center to $\sim 80\%$ in the outer part of the galaxy.

Furthermore, by identifying individual molecular clouds in M33 using {\tt CPROPS} package \citep{rosolowsky06}, we are able to measure their kinetic energy. This kinetic energy is only $\sim 1\%$ of the SNe energy integrated within the area of molecular clouds (Figure~\ref{fig:cloud_turb}). Therefore, the kinetic energy in molecular clouds can be supplied by the SNe energy. This conclusion is unaffected by the variation of CO-to-H$_2$ conversion factor due to metallicities and total surface density.

Finally, the kinetic energy from accretion can not be ruled out as a source of turbulence. From the accreted materials inferred by \citet{zheng17}, we estimate that the accreted materials have enough energy to maintain turbulence outside $\approx 4.5$ kpc radius (Figure~\ref{fig:acc_energy}). This radius may be an overestimation because the inflow rate within a galactic disk is smaller than the inferred accretion velocity of 100 km s$^{-1}$ \citep{wong04,schmidt16}, and hence, decreases the kinetic energy of inflowing materials.

\acknowledgments{
We thank the referee for improving the clarity of this manuscript. This research has been made available through the France-Berkeley Fund (FBF) as a partnership between the government of France and the University of California at Berkeley. We thank Erik Rosolowsky and Eric Koch to provide the \hi\ data from VLA and GBT, and Karl Schuster and Jonathan Braine to provide the CO map from IRAM. This work is benefited from valuable discussions with Eric Koch, Christopher McKee, Erik Rosolowsky, and Eve Ostriker. D.U. and L.B. appreciate the hospitality of \'Ecole Normale Sup\'erieure and Observatoire de Paris during their visit to Paris. D.U. and L.B. are supported by the National Science Foundation (NSF) under Grant No. AST-1140063. This research has received funding from the European Research Council under the European Community`s Seventh framework Program (FP7/2017-2022, No. 787813)

The National Radio Astronomy Observatory (NRAO) is a facility of the National Science Foundation (NSF) operated under cooperative agreement by Associated Universities, Inc. IRAM is supported by INSU/CNRS (France), MPG (Germany) and IGN (Spain). This work is based in part on observations made with the Galaxy Evolution Explorer (GALEX). GALEX is a NASA Small Explorer, whose mission was developed in cooperation with the Centre National d'Etudes Spatiales (CNES) of France and the Korean Ministry of Science and Technology. GALEX is operated for NASA by the California Institute of Technology under NASA contract NAS5-98034. This work is based in part on observations made with the Spitzer Space Telescope, which is operated by the Jet Propulsion Laboratory, California Institute of Technology under a contract with NASA. This publication makes use of data products from the Two Micron All Sky Survey, which is a joint project of the University of Massachusetts and the Infrared Processing and Analysis Center/California Institute of Technology, funded by NASA and NSF.
}

\software{MIRIAD \citep{sault95}, CPROPS \citep{rosolowsky06}, and SciPy (\url{scipy.org}).}

\facilities{VLA, GBT, IRAM, GALEX, {\it Spitzer}, and 2MASS.}

\appendix

\section{The Energy Injected by Magneto-rotational Instability} \label{app:mri}

The energy per unit area of MRI is $\Sigma_{\rm MRI} = \epsilon_{\rm MRI} \ \dot{\Sigma}_{\rm MRI} \ \tau_D$, where $0 \leq \epsilon_{\rm MRI} \leq 1$ is the coupling efficiency of MRI (i.e. the fraction of MRI energy that goes to turbulence), $\dot{\Sigma}_{\rm MRI} = 3.7 \times 10^{-8} \ {\rm erg} {\rm s}^{-1} \ {\rm cm}^{-2} \ h_{\rm \hi} \ B^2 \ S$ is the energy injection rate of MRI (described below), and $\tau_D \approx 9.8 \ {\rm Myrs} \ h_{\rm \hi} \ \sigma_{\rm HI}^{-1}$ is the dissipation time of turbulence \citep{maclow04}. Here, the units of \hi\ scale-height ($h_{\rm \hi}$), the magnetic field ($B$), the shear rate ($S$; defined in Equation~\ref{eq:shear}), and the \hi\ velocity dispersion ($\sigma_{\rm \hi}$) are 100 pc, 6$\mu$G, (220 Myrs)$^{-1}$, and 10 km s$^{-1}$, respectively. Combining it altogether, we retrieve Equation (\ref{eq:mri_final}).

The MRI energy density ($e_{\rm MRI}$) comes from the positive correlation between the radial and azimuthal components of the magnetic field (represented as the Maxwell stress tensor $T_{R \Phi}$) that transfers the energy from shear to turbulence at a rate of $\dot{e}_{\rm MRI} = T_{R \Phi} S$ \citep{sellwood99}. We adopt the value of $T_{R \Phi}$ as 0.6 times the mean magnetic energy density $B^2 (8\pi)^{-1}$ \citep{hewley95}. Then, we multiply $\dot{e}_{\rm MRI}$ by the \hi\ scale-height to get the injection rate of the MRI energy surface density, i.e. $\dot{\Sigma}_{\rm MRI} = \dot{e}_{\rm MRI} \ h_{\rm \hi}$.

\section{Rotation Curve and Shear Rate} \label{app:kin}

\begin{figure}
\centering
\includegraphics[width=\textwidth]{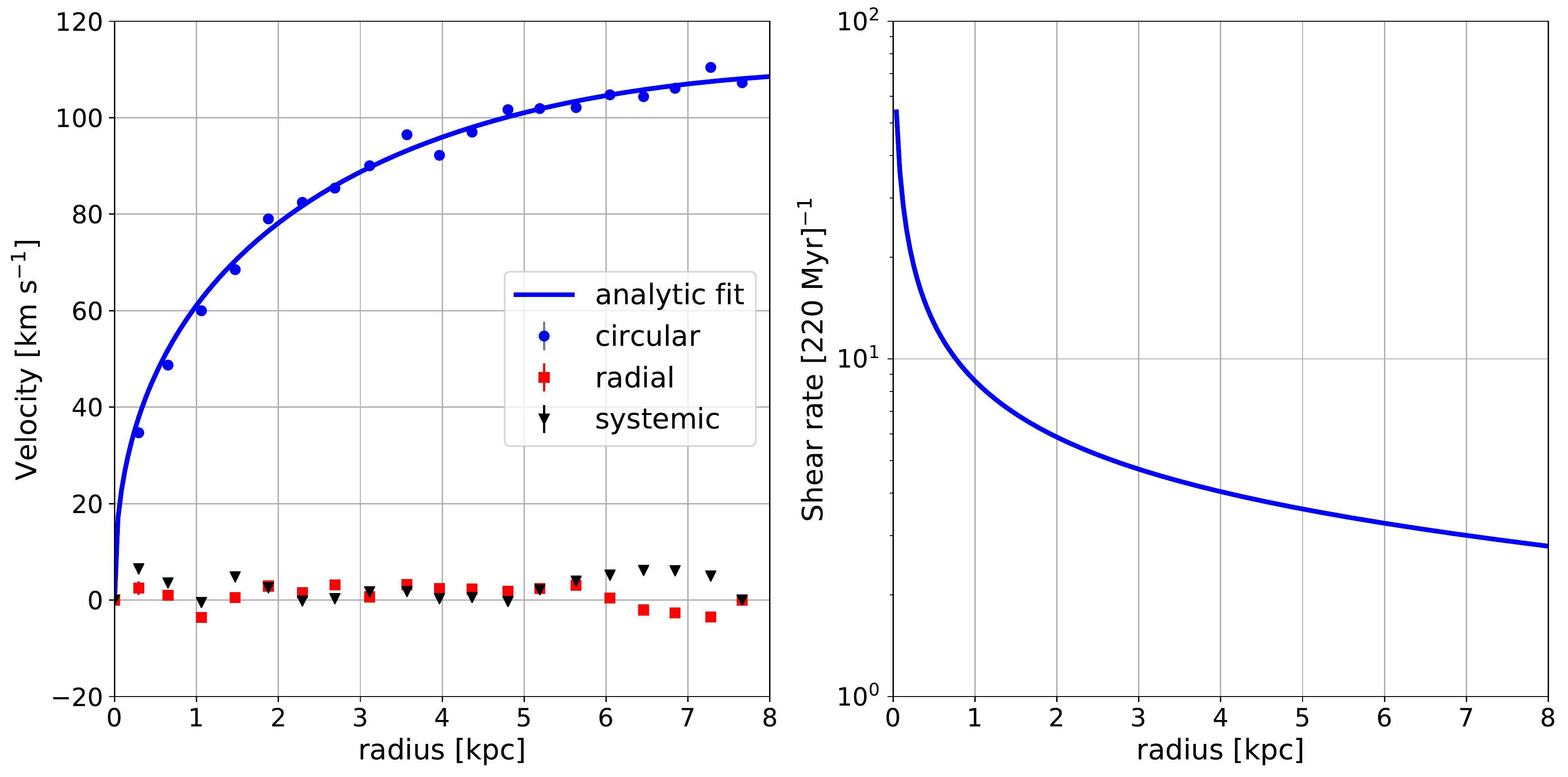}
\caption{Left: the rotation curve (blue dots), radial velocity (red squares), and systemic velocity (black triangles) of M33 as a function of radius. The uncertainties are tabulated in Table~\ref{tab:rot}. The blue curve is the analytical fit to the blue dots using Equation (\ref{eq:rotcur_fit}). Right: the shear rate as a function of radius, calculated using Equation~\ref{eq:shear}.}
\label{fig:kinematics}
\end{figure}

We derive the rotation curve from the first-moment map of atomic gas using the algorithm from \citet{bolatto02}. We fit line-of-sight velocities ($V_{\rm los}$) of each ring for the systemic ($V_{\rm sys}$), circular ($V_{\rm c}$), and radial ($V_{\rm rad}$) velocities, i.e.
\begin{equation}
V_{\rm los} = V_{\rm sys} \ + \ [V_{\rm c} \ {\rm cos}(\theta) + V_{\rm rad} \ {\rm sin}(\theta)] \ {\rm cos}(i),
\end{equation}
where $\theta$ is the angle from the kinematic major axis (receding part). In doing so we use a constant value of position angle and inclination, i.e. no warp and no isophotal twist. We fit the rotation curve using an analytical function (the blue curve on the left panel of Figure~\ref{fig:kinematics}) that takes into account the rising part as a power law and the flat part as an exponential:
\begin{equation} \label{eq:rotcur_fit}
V_c(R) = a \left( \frac{R}{R_0} \right)^b \ {\rm exp} \left( - \frac{R}{R_0} \right),
\end{equation}
where $a \approx 239.17$ km s$^{-1}$, $b \approx 0.41$, and $R_0 \approx 24.33$ kpc are the best fit parameters. Therefore the shear rate ($S$) due to differential rotation is
\begin{equation} \label{eq:shear}
S = \frac{d\Omega}{d{\rm ln}R} = V_c \left( \frac{b-1}{R} - \frac{1}{R_0} \right),
\end{equation}
and is shown on the right panel of Figure~\ref{fig:kinematics}.

\section{Tables} \label{app:table}

This section tabulates our measurements so it can be reproducible. Tables C1, C2, and C3 are published in its entirety in the machine readable format. Some portions are shown here for guidance regarding its form and content.

\begin{deluxetable*}{ c c c c c c c c c }
\label{tab:energies}
\tablewidth{0pt}
\tabletypesize{\scriptsize}
\tablecaption{The radial profile of energy per unit area}
\tablehead{
Radius & Kinetic & Thermal & Turbulent & MRI & SNe\tablenotemark{a} & SNe & Efficiency\tablenotemark{a,b} & Efficiency\tablenotemark{b} \\
kpc & $10^{46}$ erg pc$^{-2}$ & $10^{46}$ erg pc$^{-2}$ & $10^{46}$ erg pc$^{-2}$ & $10^{46}$ erg pc$^{-2}$ & $10^{46}$ erg pc$^{-2}$ & $10^{46}$ erg pc$^{-2}$ & {} & {}
}
\startdata
$0.05 \pm 0.05$ & $3.75_{-0.34}^{+0.34}$ & $0.65_{-0.25}^{+0.08}$ & $3.10_{-0.42}^{+0.59}$ & $0.18_{-0.07}^{+0.07}$ & $26.13_{-4.18}^{+3.09}$ & $32.04_{-5.13}^{+3.78}$ & $0.12_{-0.02}^{+0.03}$ & $0.10_{-0.02}^{+0.02}$ \\ 
$0.15 \pm 0.05$ & $4.12_{-1.18}^{+1.18}$ & $0.76_{-0.29}^{+0.15}$ & $3.36_{-1.33}^{+1.47}$ & $0.12_{-0.09}^{+0.12}$ & $26.62_{-7.84}^{+11.43}$ & $34.74_{-10.24}^{+14.92}$ & $0.13_{-0.06}^{+0.08}$ & $0.10_{-0.05}^{+0.06}$ \\ 
$0.25 \pm 0.05$ & $4.24_{-1.35}^{+1.35}$ & $0.83_{-0.32}^{+0.15}$ & $3.41_{-1.51}^{+1.67}$ & $0.10_{-0.08}^{+0.08}$ & $24.53_{-9.65}^{+17.56}$ & $33.76_{-13.29}^{+24.17}$ & $0.14_{-0.08}^{+0.12}$ & $0.10_{-0.06}^{+0.09}$ \\ 
$0.35 \pm 0.05$ & $4.38_{-1.57}^{+1.57}$ & $0.75_{-0.29}^{+0.18}$ & $3.63_{-1.75}^{+1.86}$ & $0.11_{-0.09}^{+0.09}$ & $15.87_{-5.09}^{+8.81}$ & $24.47_{-7.85}^{+13.58}$ & $0.23_{-0.13}^{+0.17}$ & $0.15_{-0.09}^{+0.11}$ \\ 
$0.45 \pm 0.05$ & $3.24_{-1.18}^{+1.18}$ & $0.71_{-0.28}^{+0.15}$ & $2.53_{-1.33}^{+1.46}$ & $0.09_{-0.08}^{+0.08}$ & $11.23_{-3.07}^{+3.94}$ & $18.17_{-4.96}^{+6.37}$ & $0.22_{-0.13}^{+0.15}$ & $0.14_{-0.08}^{+0.09}$ \\ 
$0.55 \pm 0.05$ & $2.75_{-0.78}^{+0.78}$ & $0.70_{-0.27}^{+0.15}$ & $2.05_{-0.94}^{+1.05}$ & $0.08_{-0.06}^{+0.07}$ & $10.52_{-3.49}^{+4.56}$ & $17.30_{-5.74}^{+7.49}$ & $0.19_{-0.11}^{+0.13}$ & $0.12_{-0.07}^{+0.08}$ \\ 
$0.65 \pm 0.05$ & $2.78_{-0.42}^{+0.42}$ & $0.71_{-0.27}^{+0.12}$ & $2.07_{-0.54}^{+0.69}$ & $0.08_{-0.04}^{+0.06}$ & $10.97_{-4.04}^{+7.86}$ & $19.11_{-7.05}^{+13.70}$ & $0.19_{-0.08}^{+0.15}$ & $0.11_{-0.05}^{+0.08}$ \\ 
$0.75 \pm 0.05$ & $2.85_{-0.55}^{+0.55}$ & $0.74_{-0.28}^{+0.13}$ & $2.11_{-0.67}^{+0.82}$ & $0.07_{-0.04}^{+0.07}$ & $15.54_{-7.25}^{+37.39}$ & $26.85_{-12.52}^{+64.60}$ & $0.14_{-0.08}^{+0.33}$ & $0.08_{-0.04}^{+0.19}$ \\ 
$0.85 \pm 0.05$ & $2.41_{-0.64}^{+0.64}$ & $0.68_{-0.26}^{+0.15}$ & $1.73_{-0.80}^{+0.90}$ & $0.07_{-0.05}^{+0.06}$ & $11.42_{-4.54}^{+13.93}$ & $20.45_{-8.13}^{+24.95}$ & $0.15_{-0.09}^{+0.20}$ & $0.08_{-0.05}^{+0.11}$ \\ 
$0.95 \pm 0.05$ & $2.19_{-0.69}^{+0.69}$ & $0.65_{-0.25}^{+0.15}$ & $1.54_{-0.84}^{+0.93}$ & $0.07_{-0.06}^{+0.07}$ & $8.67_{-3.13}^{+6.82}$ & $16.47_{-5.96}^{+12.96}$ & $0.18_{-0.12}^{+0.17}$ & $0.09_{-0.06}^{+0.09}$ \\ 
$1.05 \pm 0.05$ & $2.14_{-0.65}^{+0.65}$ & $0.63_{-0.24}^{+0.18}$ & $1.50_{-0.83}^{+0.89}$ & $0.07_{-0.06}^{+0.08}$ & $7.73_{-2.87}^{+5.71}$ & $15.42_{-5.73}^{+11.39}$ & $0.19_{-0.13}^{+0.18}$ & $0.10_{-0.06}^{+0.09}$
\enddata
\tablenotetext{a}{For a constant dissipation time of 9.8 Myr.}
\tablenotetext{b}{These are the total coupling efficiency, i.e. the turbulent energy divided by the sum of MRI and SNe energy.}
\end{deluxetable*}


\begin{deluxetable*}{ c c c c c c c }
\label{tab:rad_prof}
\tablewidth{0pt}
\tabletypesize{\scriptsize}
\tablecaption{The radial profile of surface densities, velocity dispersion, and scale-height.}
\tablehead{
Galactocentric & \multicolumn{4}{c}{Surface densities} & \hi\ Velocity & \hi\ Scale \\
radius & Atomic & Molecular\tablenotemark{a} & Stellar & SFR & dispersion & height \\
kpc & $M_\odot$ pc$^{-2}$ & $M_\odot$ pc$^{-2}$ & $M_\odot$ pc$^{-2}$ & $M_\odot$ Gyr$^{-1}$ pc$^{-2}$ & km s$^{-1}$ & pc}
\startdata
$0.05 \pm 0.05$ & $6.95_{-0.58}^{+0.59}$ & $3.57_{-1.46}^{+1.61}$ & $1018.14_{-18.58}^{+18.93}$ & $13.07_{-2.09}^{+1.54}$ & $13.46_{-0.31}^{+0.00}$ & $72.39_{-19.09}^{+20.43}$ \\ 
$0.15 \pm 0.05$ & $8.15_{-2.30}^{+2.19}$ & $5.90_{-2.83}^{+5.19}$ & $808.66_{-14.74}^{+15.01}$ & $13.31_{-3.92}^{+5.72}$ & $13.02_{-0.48}^{+0.00}$ & $74.55_{-38.52}^{+52.40}$ \\ 
$0.25 \pm 0.05$ & $8.92_{-2.73}^{+2.41}$ & $5.90_{-3.07}^{+4.04}$ & $697.71_{-12.73}^{+12.97}$ & $12.26_{-4.83}^{+8.78}$ & $12.62_{-0.80}^{+0.00}$ & $76.24_{-42.89}^{+44.67}$ \\ 
$0.35 \pm 0.05$ & $7.88_{-2.64}^{+2.31}$ & $3.53_{-1.94}^{+2.71}$ & $607.71_{-11.05}^{+11.26}$ & $7.94_{-2.55}^{+4.41}$ & $13.65_{-1.20}^{+0.08}$ & $92.31_{-57.18}^{+56.14}$ \\ 
$0.45 \pm 0.05$ & $7.52_{-2.73}^{+2.38}$ & $3.08_{-1.79}^{+2.57}$ & $539.17_{-9.84}^{+10.02}$ & $5.62_{-1.53}^{+1.97}$ & $12.01_{-0.30}^{+0.33}$ & $85.23_{-55.80}^{+55.60}$ \\ 
$0.55 \pm 0.05$ & $7.60_{-2.14}^{+1.83}$ & $3.86_{-2.38}^{+3.80}$ & $487.05_{-8.84}^{+9.01}$ & $5.26_{-1.75}^{+2.28}$ & $11.02_{-0.28}^{+0.31}$ & $79.44_{-43.74}^{+51.86}$ \\ 
$0.65 \pm 0.05$ & $7.74_{-1.13}^{+1.50}$ & $3.37_{-1.77}^{+2.70}$ & $432.19_{-3.96}^{+4.00}$ & $5.48_{-2.02}^{+3.93}$ & $10.97_{-0.26}^{+0.29}$ & $83.93_{-29.48}^{+42.29}$ \\ 
$0.75 \pm 0.05$ & $8.03_{-1.50}^{+1.59}$ & $4.77_{-2.55}^{+5.10}$ & $405.13_{-7.34}^{+7.48}$ & $7.77_{-3.62}^{+18.70}$ & $10.90_{-0.30}^{+0.38}$ & $82.64_{-35.09}^{+57.95}$ \\ 
$0.85 \pm 0.05$ & $7.46_{-1.97}^{+1.95}$ & $3.78_{-2.15}^{+3.83}$ & $390.45_{-10.58}^{+10.88}$ & $5.71_{-2.27}^{+6.97}$ & $10.40_{-0.27}^{+0.51}$ & $81.70_{-41.98}^{+55.95}$ \\ 
$0.95 \pm 0.05$ & $7.07_{-2.20}^{+2.06}$ & $2.67_{-1.64}^{+3.29}$ & $362.88_{-6.81}^{+6.94}$ & $4.33_{-1.57}^{+3.41}$ & $10.18_{-0.25}^{+0.50}$ & $84.87_{-49.73}^{+62.67}$ \\ 
$1.05 \pm 0.05$ & $6.94_{-2.09}^{+2.58}$ & $1.91_{-1.16}^{+1.95}$ & $340.21_{-6.14}^{+6.26}$ & $3.87_{-1.44}^{+2.86}$ & $10.15_{-0.26}^{+0.83}$ & $88.86_{-51.45}^{+67.99}$
\enddata
\tablenotetext{a}{Derived using a Galactic $\alpha_{\rm CO}$.}
\end{deluxetable*}


\begin{deluxetable*}{ r c c c c c c }
\label{tab:clouds}
\tablewidth{0pt}
\tabletypesize{\scriptsize}
\tablecaption{The properties of molecular clouds in M33}
\label{tab:cloud_props}
\tablehead{
No. & Radius\tablenotemark{a} & Size\tablenotemark{b} & Mass\tablenotemark{c} & Velocity Dispersion\tablenotemark{d} & Turbulent Energy & SNe energy \\
{} & kpc & pc & $10^5 M_\odot$ & km s$^{-1}$ & log($10^{51}$ erg pc$^{-2}$) & log($10^{51}$ erg pc$^{-2}$)}
\startdata
1 & $0.20 \pm 0.05$ & $89.66 \pm 16.66$ & $1.13 \pm 0.04$ & $2.57 \pm 0.91$ & $-1.60_{-0.10}^{+0.09}$ & $0.85_{-0.05}^{+0.04}$ \\ 
2 & $0.22 \pm 0.02$ & $47.00 \pm 16.66$ & $0.29 \pm 0.02$ & $2.82 \pm 1.10$ & $-2.13_{-0.19}^{+0.17}$ & $0.07_{-0.03}^{+0.03}$ \\ 
3 & $0.24 \pm 0.05$ & $42.04 \pm 16.66$ & $0.50 \pm 0.02$ & $3.89 \pm 1.31$ & $-1.63_{-0.16}^{+0.15}$ & $0.64_{-0.04}^{+0.04}$ \\ 
4 & $0.39 \pm 0.11$ & $99.03 \pm 16.66$ & $1.18 \pm 0.03$ & $2.81 \pm 1.68$ & $-1.39_{-0.14}^{+0.12}$ & $0.82_{-0.04}^{+0.03}$ \\ 
5 & $0.55 \pm 0.03$ & $79.76 \pm 16.66$ & $1.17 \pm 0.05$ & $3.14 \pm 0.91$ & $-1.41_{-0.12}^{+0.10}$ & $0.32_{-0.04}^{+0.04}$ \\ 
6 & $0.55 \pm 0.12$ & $87.17 \pm 16.66$ & $1.11 \pm 0.05$ & $2.65 \pm 1.13$ & $-1.55_{-0.12}^{+0.10}$ & $0.51_{-0.02}^{+0.02}$ \\ 
7 & $0.69 \pm 0.03$ & $54.00 \pm 16.66$ & $0.56 \pm 0.02$ & $3.76 \pm 1.33$ & $-1.58_{-0.13}^{+0.10}$ & $-0.35_{-0.01}^{+0.01}$ \\ 
8 & $0.75 \pm 0.04$ & $59.45 \pm 16.66$ & $0.71 \pm 0.04$ & $2.61 \pm 0.51$ & $-1.82_{-0.09}^{+0.09}$ & $0.09_{-0.03}^{+0.03}$ \\ 
9 & $0.76 \pm 0.05$ & $63.05 \pm 16.66$ & $1.69 \pm 0.13$ & $3.61 \pm 0.55$ & $-1.17_{-0.10}^{+0.08}$ & $1.32_{-0.10}^{+0.09}$ \\ 
10 & $0.76 \pm 0.03$ & $49.74 \pm 16.66$ & $0.55 \pm 0.04$ & $2.71 \pm 0.43$ & $-1.91_{-0.11}^{+0.09}$ & $0.33_{-0.05}^{+0.05}$
\enddata
\tablenotetext{a}{Distance is measured from the nucleus of M33.}
\tablenotetext{b}{Size is defined as $({\rm area}/\pi)^{0.5}$. The uncertainty is the physical size of one pixel.}
\tablenotetext{c}{Derived using a variable $\alpha_{\rm CO}$. The uncertainty is calculated using bootstrap resampling with 1,000 iterations.}
\tablenotetext{d}{The mean velocity dispersion within a cloud. The uncertainty is standard deviation of velocity dispersion within a cloud.}
\end{deluxetable*}


\begin{deluxetable*}{ c c c c c c c c c }
\label{tab:rot_cur}
\tablewidth{0pt}
\tabletypesize{\scriptsize}
\tablecaption{The rotation curve of M33}
\label{tab:rot}
\tablehead{
Radius & Circular velocity & Radial velocity & Systemic velocity \\
kpc & km s$^{-1}$ & km s$^{-1}$ & km s$^{-1}$}
\startdata
$0.3$ & \hspace{2mm}$34.7 \pm 0.9$ & \hspace{3mm}$2.5 \pm 1.4$ & \hspace{3mm}$6.5 \pm 1.1$ \\
$0.7$ & \hspace{2mm}$48.7 \pm 0.3$ & \hspace{3mm}$1.0 \pm 0.4$ & \hspace{3mm}$3.6 \pm 0.3$ \\
$1.1$ & \hspace{2mm}$60.0 \pm 0.2$ & \hspace{3mm}$-3.6 \pm 0.2$ & \hspace{3mm}$-0.5 \pm 0.2$ \\
$1.5$ & \hspace{2mm}$68.5 \pm 0.2$ & \hspace{3mm}$0.5 \pm 0.2$ & \hspace{3mm}$4.8 \pm 0.1$ \\
$1.9$ & \hspace{2mm}$79.0 \pm 0.1$ & \hspace{3mm}$2.9 \pm 0.1$ & \hspace{3mm}$2.6 \pm 0.1$ \\
$2.3$ & \hspace{2mm}$82.5 \pm 0.1$ & \hspace{3mm}$1.5 \pm 0.1$ & \hspace{3mm}$-0.2 \pm 0.1$ \\
$2.7$ & \hspace{2mm}$85.4 \pm 0.1$ & \hspace{3mm}$3.1 \pm 0.1$ & \hspace{3mm}$0.3 \pm 0.1$ \\
$3.1$ & \hspace{2mm}$90.0 \pm 0.1$ & \hspace{3mm}$0.7 \pm 0.1$ & \hspace{3mm}$1.7 \pm 0.1$ \\
$3.6$ & \hspace{2mm}$96.5 \pm 0.1$ & \hspace{3mm}$3.2 \pm 0.1$ & \hspace{3mm}$1.8 \pm 0.1$ \\
$4.0$ & \hspace{2mm}$92.2 \pm 0.1$ & \hspace{3mm}$2.4 \pm 0.1$ & \hspace{3mm}$0.3 \pm 0.1$ \\
$4.4$ & \hspace{2mm}$97.0 \pm 0.1$ & \hspace{3mm}$2.3 \pm 0.1$ & \hspace{3mm}$0.6 \pm 0.1$ \\
$4.8$ & $101.7 \pm 0.1$ & \hspace{3mm}$1.9 \pm 0.1$ & \hspace{3mm}$-0.3 \pm 0.1$ \\
$5.2$ & $101.9 \pm 0.1$ & \hspace{3mm}$2.4 \pm 0.1$ & \hspace{3mm}$2.1 \pm 0.1$ \\
$5.6$ & $102.1 \pm 0.1$ & \hspace{3mm}$3.0 \pm 0.1$ & \hspace{3mm}$4.0 \pm 0.1$ \\
$6.0$ & $104.8 \pm 0.1$ & \hspace{3mm}$0.4 \pm 0.1$ & \hspace{3mm}$5.2 \pm 0.1$ \\
$6.5$ & $104.4 \pm 0.1$ & \hspace{3mm}$-2.0 \pm 0.1$ & \hspace{3mm}$6.2 \pm 0.0$ \\
$6.8$ & $106.1 \pm 0.1$ & \hspace{3mm}$-2.7 \pm 0.1$ & \hspace{3mm}$6.1 \pm 0.1$ \\
$7.3$ & $110.4 \pm 0.1$ & \hspace{3mm}$-3.5 \pm 0.1$ & \hspace{3mm}$5.0 \pm 0.1$ \\
$7.7$ & $107.3 \pm 0.5$ & \hspace{3mm}$0.0 \pm 0.0$ & \hspace{3mm}$0.0 \pm 0.0$ \\
$8.1$ & $111.6 \pm 0.9$ & \hspace{3mm}$0.0 \pm 0.0$ & \hspace{3mm}$0.0 \pm 0.0$
\enddata
\end{deluxetable*}

\section{The Energy Injected by Supernovae} \label{app:sne}

{\it Supernovae Rates per Unit Area ($\eta$):} Following \citet{tamburro09}, the rate of supernovae explosions is given by the average number of newly formed stars (SFR$/\bar{m}_*$) multiplied by the fraction of the newly formed stars that become supernovae ($f_{\rm SN}$). Here, $\bar{m}_*$ is the average mass of a stellar population. If we assume IMF as $\phi(m) \propto m^{-\alpha}$, where $\alpha = 1.3$ for $0.1 < M/M_{\odot} < 0.5$ and $\alpha = 2.3$ for $0.5 < M/M_{\odot} < 120$ \citep{calzetti07}, then $\bar{m}_*$ and $f_{\rm SN}$ are given by
\begin{equation} \label{eq:m_star}
\bar{m}_* = \frac{\int_{0.1 M_{\odot}}^{120 M_{\odot}} m \ \phi(m) \ dm}{\int_{0.1 M_{\odot}}^{120 M_{\odot}} \phi(m) \ dm},
\end{equation}
\begin{equation} \label{eq:f_sn}
f_{\rm SN} = \frac{\int_{8 M_{\odot}}^{120 M_{\odot}} \phi(m) \ dm}{\int_{0.1 M_{\odot}}^{120 M_{\odot}} \phi(m) \ dm}.
\end{equation}
\noindent In Equations (\ref{eq:m_star}) and (\ref{eq:f_sn}) we assume that only stars with $8 < M/M_{\odot} < 120$ go into SNe. Therefore, for core collapse (Type Ia) SNe,
\begin{equation} \label{eq:eta}
\eta = \frac{f_{\rm SN}}{\bar{m}_*} \ \Sigma_{\rm SFR} \approx 1.3 \times 10^{-5} \ {\rm yr}^{-1} \ {\rm kpc}^{-2} \left( \frac{\Sigma_{\rm SFR}}{M_{\odot} \ {\rm Gyr}^{-1} \ {\rm pc}^{-2}} \right).
\end{equation}
Note that \citet{mannucci05} found that the rate of Type Ia SNe is few times lower than Type II for Sb-c type galaxy.

{\it Dissipation Time Scale ($\tau_D$):} Since the energy dissipation happens at the cooling radius ($R_C$), we can estimate the driving scale  for inhomogeneous medium as \citep{martizzi15}
\begin{equation}
R_C \approx 6.3 \ {\rm pc} \ \left( \frac{n_H}{100 \ {\rm cm}^{-3}} \right)^{-0.42} \approx 43.6 \ {\rm pc \ for} \ n_H = 1 \ {\rm cm}^{-3}.
\end{equation}
Note that from simulations by \citet{padoan16}, the scale where most of kinetic energy (injected by SNe) is contained is $\approx 70$ pc. This $R_C$ gives a dissipation time as \citep{maclow04}
\begin{equation} \label{eq:tau_d}
\tau_D \approx 9.8 \ {\rm Myr} \ \left( \frac{R_C}{\rm 100 \ pc} \right) \ \left( \frac{\sigma}{\rm 10 \ km/s} \right)^{-1} \approx 4.27 \ {\rm Myr}.
\end{equation}

{\it Energy Injected by Single Supernovae ($E_{\rm SN}$):} To calculate the momentum injected by a SN ($p_*$), we use the value from the simulation of \citet{kim15} for a single SN in two-phase medium as
\begin{equation}
p_* \approx 2.8 \times 10^5 \ M_{\odot} \ {\rm km \ s}^{-1} \ \left( \frac{n_H}{100 \ {\rm cm}^{-3}} \right)^{-0.17}.
\end{equation}
For $n_H = 1$ cm$^{-3}$, the momentum injected by a SN is $p_* \approx 1.2 \times 10^{44}$ g cm s$^{-1}$. This momentum sweeps out and injects energy to ISM at $R_C$. The mass of this ISM is about
\begin{equation}
M_{\rm ISM} = \frac{4 \pi}{3} \ R_C^3 \ \rho_{\rm ISM} \approx 2 \times 10^{37} \ {\rm gram},
\end{equation}
for $\rho_{\rm ISM} = 2 \times 10^{-24}$ g cm$^{-3}$. Therefore the energy injected by a SN into ISM is
\begin{equation} \label{eq:E_sn}
E_{\rm SN} = \frac{p_*^2}{2 M_{\rm ISM}} \approx 3.6 \times 10^{50} \ {\rm erg},
\end{equation}
which is 3 times lower than the common assumption of SN energy of $10^{51}$ erg. This is due to the fact that not all SN energy goes into ISM kinetic energy.

{\it Total Energy Injected by Supernovae:} Combining Equations (\ref{eq:eta}), (\ref{eq:tau_d}), and (\ref{eq:E_sn}), we get
\begin{equation}
\Sigma_{\rm SNE} = \eta \ \epsilon_{SN} \ E_{\rm SN} \ \tau_D \approx \Sigma_{\rm SNE,0} \  \epsilon_{SN} \ \left( \frac{\Sigma_{\rm SFR}}{M_{\odot} \ {\rm Gyr}^{-1} \ {\rm pc}^{-2}} \right),
\end{equation}
where $\Sigma_{\rm SNE,0} \approx 2.0 \times 10^{46} \ {\rm erg \ pc}^{-2}$ and $0 \leq \epsilon_{SN} \leq 1$.

\section{The Effect of Beam Smearing to the Measured Velocity Dispersion}
\label{app:smear}

For a finite size of the observing beam, the rotation of the gas from a galaxy with non-zero inclination (not face-on) would create an artificial broadening of the spectral line \citep[see e.g.,][]{federrath16,federrath17,leung18,levy18,sharda18}. This is because rotating ``particles'' at different projected locations inside the beam have different line-of-sight velocities. This effect would make the velocity dispersion ($\sigma_{\hi}$) appear larger than what it should be. Investigating this artificial broadening is important because the gas turbulent energy depends on $\sigma_{\hi}^2$.

\begin{figure*}
\epsscale{1.15}
\plotone{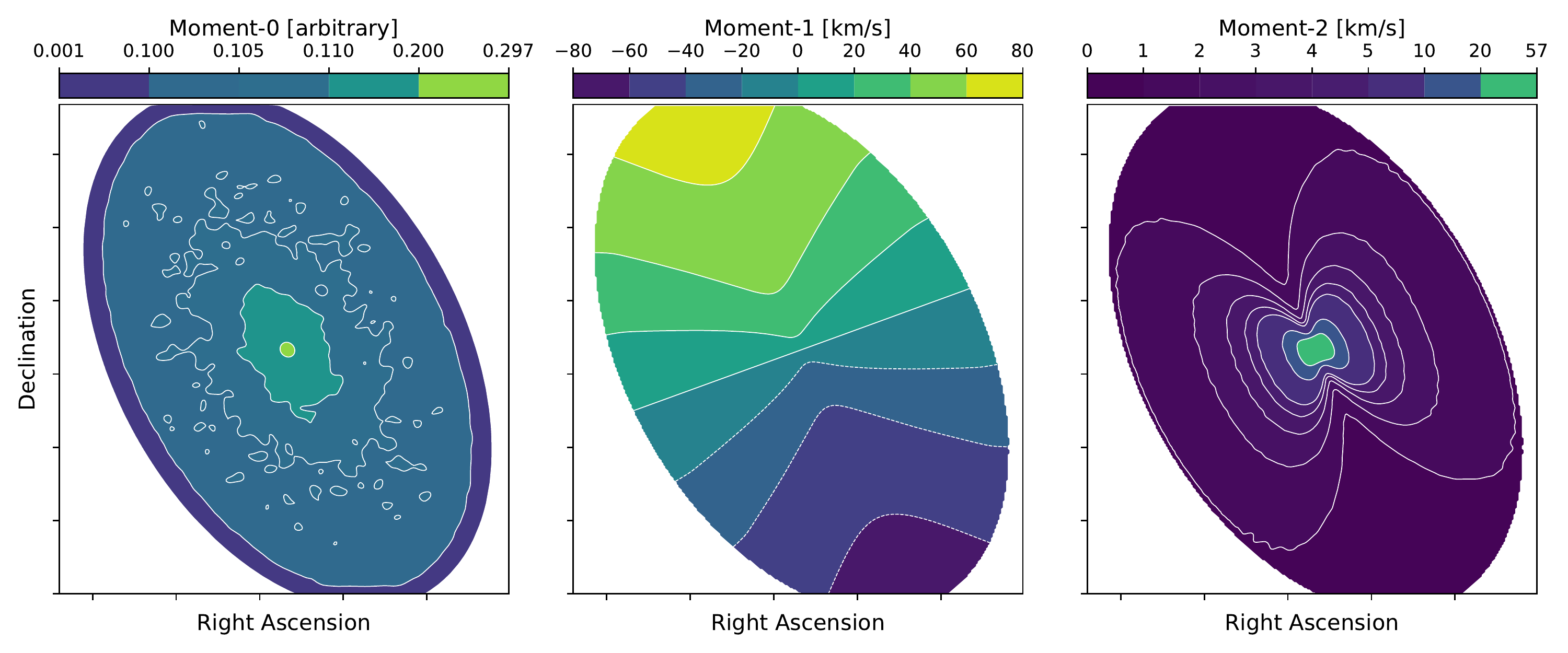}
\plottwo{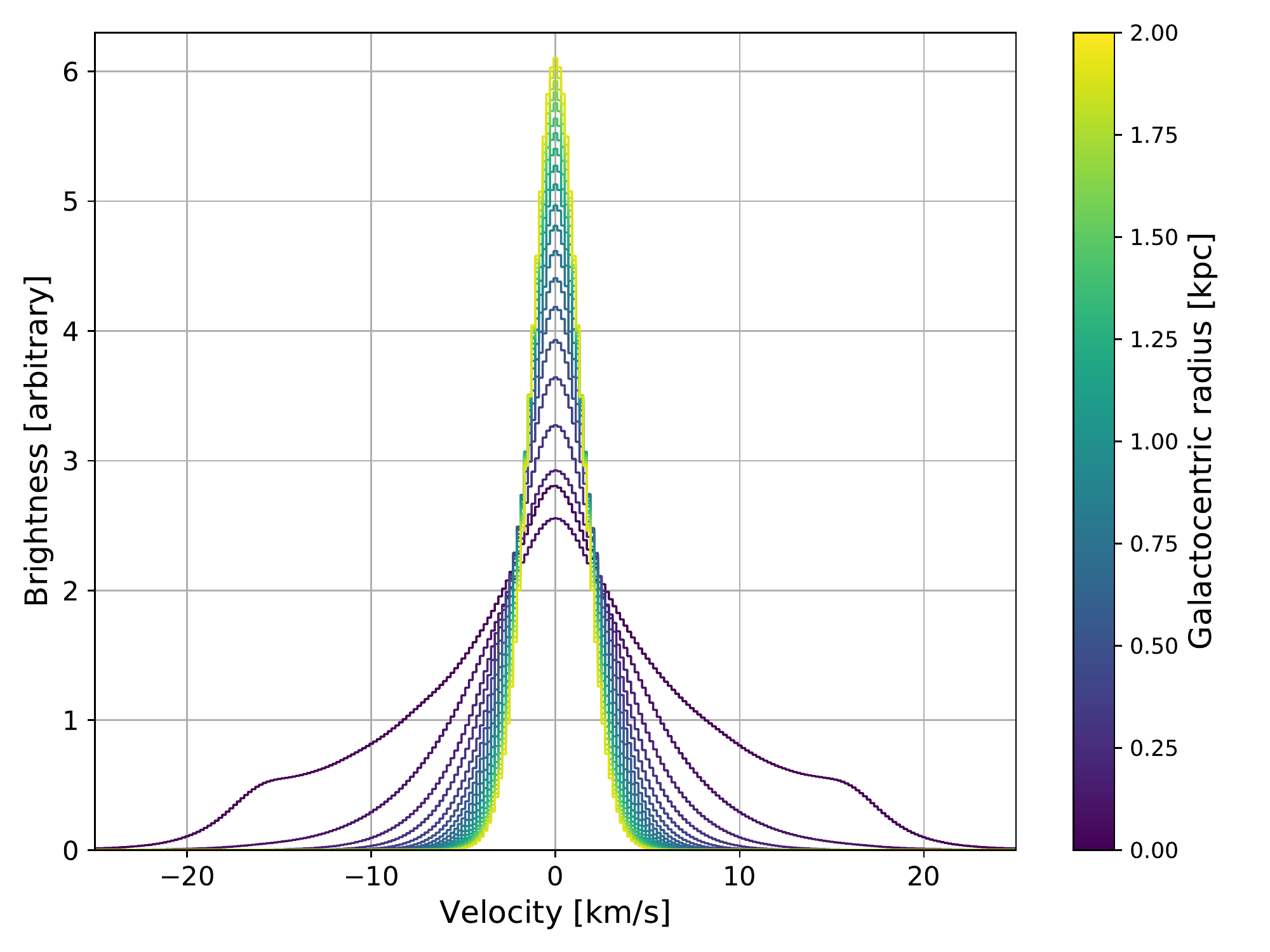}{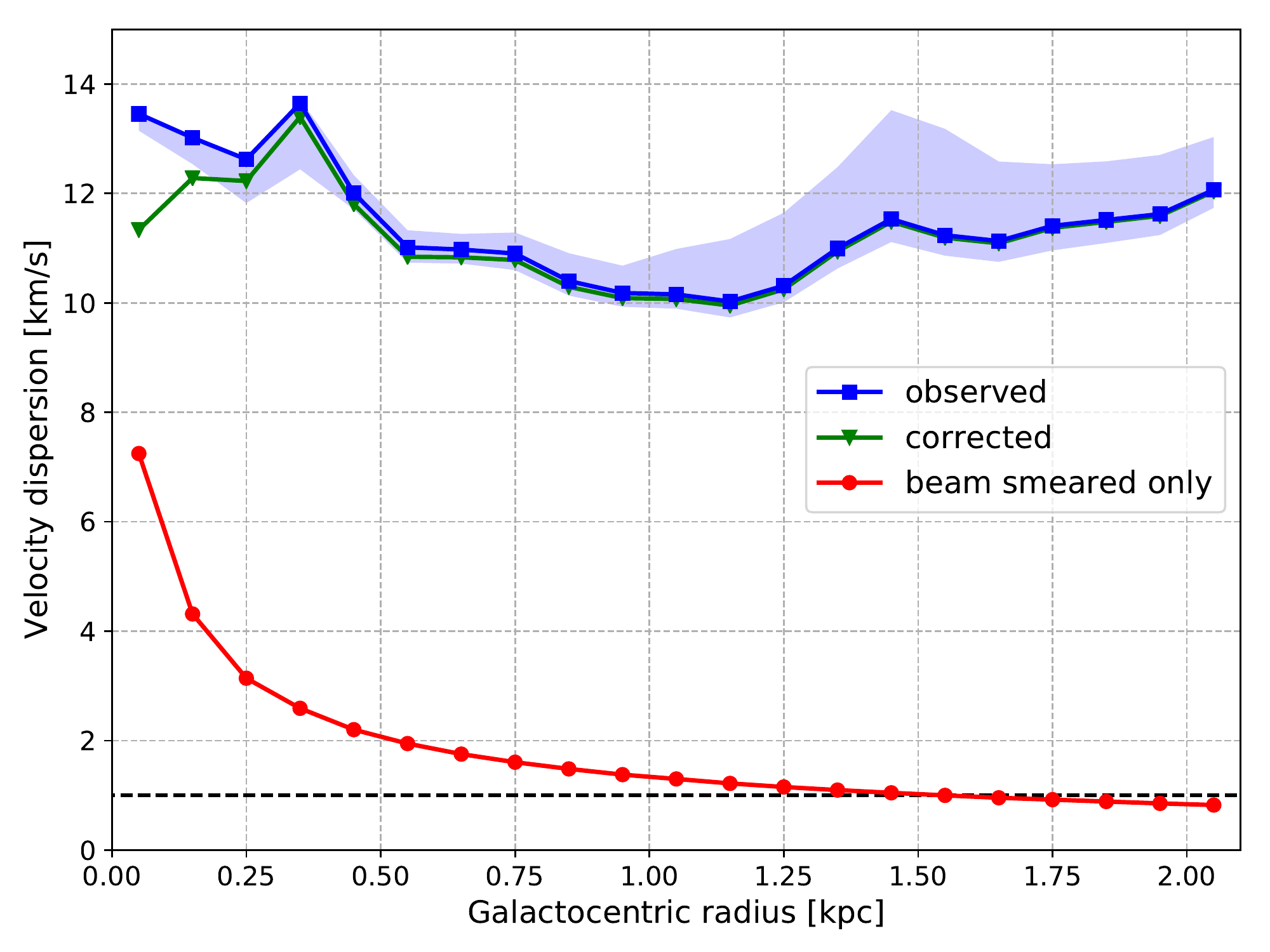}
\caption{Outcomes of simulation using the Kinematic Molecular Simulation (KinMS) package of \citet{davis13b,davis13a} to mimic the rotation in M33 inside 2~kpc radius. {\it Top panels:} simulated moment-0 (left), moment-1 (middle), and moment-2 (right). This shows that the pure rotation causes an artificial moment-2 due to beam smearing. {\it Bottom left panel:} stacked spectra within annulus ring with 100~pc width. The color shows the distance of each ring from the center. The center shows a significant broadening with dispersion of $\approx7.25$~km~s$^{-1}$, measured from a Gaussian fit. {\it Bottom right panel:} the velocity dispersion due to beam smearing (red) and the observed velocity (blue) as a function of distance from the center. The difference between those two dispersions is shown as the corrected velocity dispersion (green).}
\label{fig:smear}
\end{figure*}

Here, we measure the amount of this artificial broadening due to beam smearing by creating a simulated galaxy with rotation, inclination, position angle, and the thickness of gas disk equivalent to those in M33, but with zero intrinsic velocity dispersion. To do so, we use the Kinematic Molecular Simulation (KinMS) package of \citet{davis13b,davis13a}. We set the physical resolution of this simulated galaxy to be 1\arcsec\ (a factor of 20 smaller than the observed beam size), the same velocity resolution as the observed one (0.2~km~s$^{-1}$), and $10^7$ ``cloudlets'' that spread over the simulated area to create a smooth emission. To reduce the computing time, we only do the simulation for the inner $\approx 2$~kpc from the center. The beam smearing effect is prominent in the center, where the rotation curve is rising, and becomes almost negligible outward, where the rotation curve becomes flatter. Then, we convolve and regrid the simulated cube to match the observed resolution and pixel scale. We show the zeroth, first, and second moments maps of the simulated cube in the top panels of Figure~\ref{fig:smear}. The contour shape in the second moment is expected because the beam smearing effect is larger in regions where the contours of equal line-of-sight velocity (as shown in the first moment map) is closer to one another.

As in the observation, we stack the simulated spectra within 100~pc radial bins. We plot this stacked spectra in the bottom left panel of Figure~\ref{fig:smear}. As expected, the center has a significant artificial broadening, and it gets smaller for farther radial distance from the center. We fit this stacked spectra with a Gaussian to measure its velocity dispersion, shown as red dots in the right panel of Figure~\ref{fig:smear}. As a comparison, the observed velocity dispersion (blue) is also shown. From 1.5~kpc outwards, the dispersion from beam smearing is $<1$~km~s$^{-1}$ (or $<10\%$ of the observed dispersion) and the trend is flattening. The dashed horizontal line marks 1~km~s$^{-1}$ of velocity dispersion. Inside 0.5~kpc, the dispersion from beam smearing increases rapidly from 2~km~s$^{-1}$ to 7.25~km~s$^{-1}$ in the center.

We express the square of the corrected velocity dispersion ($\sigma_{\rm cor}$) as quadrature difference between the observed ($\sigma_{\rm obs}$) and the beam smeared velocity dispersion ($\sigma_{\rm beam}$),
\begin{equation}
\sigma_{\rm cor} = \sqrt{\sigma_{\rm obs}^2 - \sigma_{\rm beam}^2}~.
\end{equation}
We plot this $\sigma_{\rm cor}$ as green curve in Figure~\ref{fig:smear}. {\it Thus, beam smearing is not an issue outside 250~pc from the center.} In the center, $\sigma_{\rm obs}$ is larger than $\sigma_{\rm cor}$ by about $19\%$. Since the gas kinetic energy is proportional to the square of velocity dispersion, this means we overestimate the kinetic energy in the center by $\approx42\%$.

\bibliographystyle{yahapj}
\bibliography{references}

\end{document}